\pdfoutput=1
\documentclass[12pt]{article}

.5 scaled \magstep4
.5 scaled \magstep2
\outer\def\beginsection#1\par{\medbreak\bigskip
      \message{#1}\leftline{\bf#1}\nobreak\medskip
\vskip-\parskip
      \noindent}
\usepackage[scr=rsfs,cal=boondox]{mathalfa}
\usepackage[overload]{textcase}
\usepackage{graphicx} 
\begin{document}
\bibliographystyle{unsrt}

\begin{center}
{\Large {\bf Fuzzy bounces}}\\
\vspace{15mm}
 Massimo Giovannini
 \footnote{Electronic address: massimo.giovannini@cern.ch} \\
\vspace{1cm}
{{\sl  Department of Physics, CERN CH-1211 Gen\`eve 23, Switzerland }}
\vspace{0.5cm}\\
{{\sl INFN, Section of Milan-Bicocca, 20126 Milan, Italy}}
\vspace*{1cm}
\end{center}
We observe that the energy and the enthalpy densities can be smeared by 
two fudge factors that are constrained by the contracted Bianchi identities. Depending on the analytic properties of the smearing functions the underlying cosmological solutions belong to two physically different classes, namely the bounces of the scale factor and the curvature bounces. While the curvature bounces are naturally compatible with a stage of accelerated expansion, the bounces of the scale factor demand an early phase of accelerated contraction even if a short inflationary stage may arise prior to the decelerated regime. Despite the regularity of the underlying solutions, gradient instabilities and singularities do occasionally appear in the evolution of curvature inhomogeneities. After deducing the specific criteria behind these occurrences, the background-independent conclusions are corroborated by a series of concrete examples associated with different forms of the smearing functions. The evolution of the curvature inhomogeneities restricts the ranges of the solutions that turn out to be unsuitable even for a limited description of the pre-inflationary initial data.  The same observation holds in the case of the gauge-invariant evolution of the matter density contrast. It is however not excluded that a class of scenarios (mainly associated with the curvature bounces) could indeed avoid the potential instabilities. All in all the present analysis explore a general approach whose results are relevant in all the contexts where bouncing solutions are invoked either as complementary or as alternative to the conventional inflationary scenarios. 
\vskip 0.5cm

\nonumber
\noindent

\vspace{5mm}

\vfill
\newpage

\newpage
\renewcommand{\theequation}{1.\arabic{equation}}
\setcounter{equation}{0}
\section{Formulation of the problem}
\label{sec1}
In the last score year the upper limits on the tensor to scalar ratio ($r_{T}$ in what follows) became progressively more stringent: from the first few WMAP data releases \cite{WW1,WW2a,WW2b} the allowed $r_{T}$ experienced a steady reduction so that the direct bounds now require that $r_{T} < {\mathcal O}(10^{-2})$ \cite{WW3,WW4} at a typical pivot scale $k_{p} = 0.002 \, \mathrm{Mpc}^{-1}$. Even if the current bounds on $r_{T}$ restrict the classes of the allowed potentials, the conventional inflationary scenarios \cite{WW5,WW5a,WW5b,WW5c} (see also \cite{WW6}) 
seem still consistent with the so-called adiabatic paradigm \cite{ADP1,ADP2} where a single adiabatic mode dominates the initial conditions of the temperature and polarization anisotropies of the cosmic microwave background.  While a plausible 
class of models involves the so-called plateau-like potentials \cite{WW5} (see also \cite{WW7,WW7a}), the resulting scenarios must always be appropriately fine-tuned. To some the nature of these tunings appears controversial since inflation is supposed to produce the observable Universe from a generic set of initial data and this means that, for sufficiently large length-scales (e.g. $\ell \,> 10^{6}\, \ell_{P}$) before the onset  of the inflationary stage, the potential and the kinetic energies of the inflaton should be comparable (say, within one order of magnitude) with the contributions of the spatial gradients and of the intrinsic curvature. Whenever the (approximate) equipartition between the different components is violated, the potential at the onset of inflation might undershoot the other components of the pre-inflationary plasma. If not adequately tuned, the intrinsic curvature (or the kinetic energy) might dominate the evolution prior to the dominance of the potential; the Universe may then get inhomogeneous (or simply decelerated) prior to the onset of the inflationary stage of expansion\footnote{While 
this viewpoint can be quantitatively scrutinized by following the evolution of the spatial gradients 
during the preinflationary phase \cite{WW8,WW8a,WW8b},  a satisfactory theory of the initial conditions should probably account for the current expanding stage of the Universe from a reasonably generic set of initial data, as repeatedly propounded, for instance, in Refs. \cite{WW7,WW7a}. If this does not happen, inflation should be primarily regarded 
as a model of the spectral indices rather than a theory of the initial data.}.  

According to current data the energy density associated with plateau-like potentials prior to the onset of inflation should be smaller than $10^{-12}$ in Planckian units \cite{WW8b}. To be fair the possibility of a negligible tensor to scalar ratio (i.e. $r_{T} \ll 1$) can also be realized when the spectrum of the relic gravitons is strongly peaked at high-frequencies \cite{MGinv} so that the low- frequency gravitons remain invisible in the aHz region but are still potentially detectable at much higher frequencies. In this framework $r_{T}$ could be as small as ${\mathcal O}(10^{-7})$  in the aHz region but still potentially detectable in the MHz range  \cite{MGinv} (see also \cite{MGinv2}).  Along a complementary perspective it has been argued that for hilltop potentials $r_{T}$ can be arbitrarily reduced \cite{HT1} so that models with parametrically small $r_{T}$ can be constructed in specific situations \cite{HT2}. These examples may arise in certain classes of fast-roll potentials (see, for instance, \cite{FR1,FR2,FR3}). 

A less conservative option leading to a sharp suppression of $r_{T}$ involves 
bouncing models that may lead to $r_{T} \ll {\mathcal O}(10^{-2})$ over large 
distance scales \cite{RELGRAV} corresponding to typical frequencies of the order of $3$ aHz (recall $1\, \mathrm{aHz} = 10^{-18} \mathrm{Hz}$). These scenarios have been intensively investigated in the last thirty years with various independent motivations ranging from string inspired models to different classes of models concocted in the framework of effective theories (see, e.g. \cite{BB1,BB2,BB3,BB4}).
A stage of accelerated contraction goes back to the pioneering ideas of Tolman and Lema\^itre   \cite{OR1,OR2} even if both in these initial attempts and in some of the subsequent developments \cite{OR3,OR4,OR5} the spatial (intrinsic) curvature played a crucial role. However both a stage of accelerated contraction as well as a phase of accelerated 
expansion may dilute the spatial gradients and suppress the spatial curvature \cite{BB1,BB2,BB3,DD1,DD2, BB4} and, for this reason, a bouncing stage is either propounded as a viable alternative to a stage of accelerated expansion or as 
an early-time completion of a conventional inflationary scenario.

 On a more physical ground, in what follows we shall often refer to the distinction between the  bounces of the scale factor and the curvature bounces. The extrinsic (Hubble) curvature locally vanishes for the bounces of the scale factor while in the case of the  curvature bounces it is the time derivative of the Hubble rate that goes to zero. Both possibilities will then be considered here in the case of vanishing intrinsic (spatial) curvature which is always dynamically suppressed in both situations. Let us then start from the standard form of the gravity action supplemented by the matter sources in four-dimensional space time\footnote{For the sake of accuracy we mention that the signature of the metric is 
mostly minus (i.e. $[+,\,-,\,-,\,-]$); the Greek indices are four-dimensional while Latin (lowercase) indices 
are purely spatial.}:
\begin{equation}
S_{tot} = - \frac{1}{2 \, \ell_{P}^2} \int \, d^{4} x\, \sqrt{- g} \, R + S_{m}, \qquad \qquad \ell_{P} = \sqrt{8\pi\,G} = 1/\overline{M}_{P},
\label{EQ1}
\end{equation}
where $g$ denotes the determinant of the four-dimensional metric, $\ell_{P} = \sqrt{8 \, \pi\, G}$ and $R$ is the Ricci scalar. In Eq. (\ref{EQ1}) $S_{m}$ indicates the matter action that can assume different forms; to illustrate the idea pursued here we therefore consider, for the sake 
of concreteness, the case of a perfect irrotational fluid where the corresponding energy-momentum tensor can be written as:
\begin{equation}
{\mathcal t}_{\mu}^{\,\,\nu} = - (p + \rho) \,\, {\mathcal P}_{\mu}^{\,\,\nu} + \rho\,\, \delta_{\mu}^{\,\,\nu},
\label{EQ2}
\end{equation}
where $\rho$ denotes throughout the energy density while $(p\, + \rho)$ is the enthalpy density. Furthermore we recall that
${\mathcal P}_{\mu}^{\,\, \nu}$ is the standard covariant projector obeying:
\begin{equation}
{\mathcal P}_{\mu}^{\,\, \nu} = \bigl(\delta_{\mu}^{\,\,\nu} - u_{\mu} \, u^{\, \nu}\bigr), \qquad u_{\nu} \, {\mathcal P}^{\mu\nu} =0, \qquad g_{\mu\nu} \,{\mathcal P}^{\mu \nu} = 3.
\label{EQ2a}
\end{equation}
From Eqs. (\ref{EQ2})--(\ref{EQ2a}) the explicit forms of the energy density and of the pressure follow by projecting ${\mathcal t}^{\mu\nu}$ along $u_{\mu}\, u_{\nu}$ and along ${\mathcal P}_{\mu\nu}$. The standard 
form  of Eq. (\ref{EQ2}) can always be associated with a modified energy-momentum tensor involving 
two different smearing functions associated, respectively, with the energy and with the enthalpy densities: 
\begin{equation}
{\mathcal t}_{\mu}^{\,\,\nu}(\rho, p) \, \qquad \Rightarrow \qquad T_{\mu}^{\,\, \nu}(\rho,\, p) = - (p+ \rho) \, q(\rho)\, {\mathcal P}_{\mu}^{\,\,\nu} + 
\rho\, f(\rho) \, \delta_{\mu}^{\,\, \nu}.
\label{EQ3}
\end{equation}
While more general possibilites can be imagined, in  Eq. (\ref{EQ3}) $q(\rho)$ and $f(\rho)$ correspond to the two fudge factors that  solely depend upon the energy density of the matter fields. With equivalent terminology $q(\rho)$ and $f(\rho)$ are two appropriate smearing functions whose properties, as we shall see in section \ref{sec2}, are constrained by the contracted Bianchi identities.
The smearing functions determine not only the features of the bouncing solutions but also the evolution and the potential singularities of the curvature inhomogeneities.

All in all the layout of this paper is, in short, the following. In section \ref{sec2} the general aspects on the parametrization introduced in Eq. (\ref{EQ3}) are scrutinized with particular attention to the relevant constraints 
implied by the covariant conservation of the transformed energy-momentum tensor. 
The cases that are compatible with the isotropy of the background are preferentially analyzed and, for this reason, 
the ${\mathcal t}_{\mu}^{\nu}$ will either be associated 
with a perfect irrotational fluid or with a single scalar field; towards the end of section \ref{sec2} some possible extensions involving the viscous fluids will be outlined. The evolution of the curvature inhomogeneities in the presence of a bouncing dynamics is analyzed in section \ref{sec3} without specifying the 
details of the solutions. By using the physical properties of the smearing functions we shall be able to deduce the general properties of the curvature inhomogeneities and the potential drawbacks associated 
with their evolution. The results of section \ref{sec3} are corroborated by a series of explicit examples 
that are discussed in section \ref{sec4}. Section \ref{sec5} contains our concluding remarks. The relevant 
details associated with the evolution of the curvature inhomogeneities and with the explicit 
forms of the pump fields in different time parametrizations have been relegated, respectively, 
to appendices \ref{APPA} and \ref{APPB}.  In appendix \ref{APPC} the interested reader may find a dedicated discussion 
on the evolution of the gauge-invariant density contrast. As a general comment we stress that 
the cases associated with the bounces of the scale factor and of the curvature have been 
explicitly separated in dedicated subsections; this we did not for sake of pedantry but just 
to keep separated, even formally, the different dynamical situations.

\renewcommand{\theequation}{2.\arabic{equation}}
\setcounter{equation}{0}
\section{General constraints on the parametrization}
\label{sec2}
\subsection{Perfect irrotational fluids}
When the energy-momentum tensor $T_{\mu\nu}$ appearing in Eq. (\ref{EQ3}) is the source of the Einstein's equation it must be covariantly conserved because of the contracted Bianchi identities; in other words we should have that
\begin{equation}
{\mathcal G}_{\mu}^{\,\,\nu} = \ell_{P}^2 \,\,T_{\mu}^{\,\,\nu}, \qquad \qquad \nabla_{\mu} \,T^{\mu\nu} =0.
\label{CC0b}
\end{equation}
In general the covariant conservation of Eq. (\ref{CC0b}) does not constrain directly the two smearing functions $f(\rho)$ and $q(\rho)$. However we may require that the transformation between ${\mathcal t}_{\mu}^{\,\,\nu}$ and $T_{\mu}^{\,\,\nu}$ preserves the original form of the covariant conservation. In other words we shall impose that  
$\nabla_{\mu} \,T^{\mu\nu} =0$ implies the covariant conservation of $ {\mathcal t}^{\mu\nu}$ and vice versa. With this requirement $q(\rho)$ and $f(\rho)$ do affect the corresponding Einstein's equations at the price of a specific relation between the two smearing functions. The conditions coming from the covariant conservation follow by projecting $\nabla_{\mu}\,T^{\mu\nu}$
along $u_{\nu}$ and in the orthogonal direction ${\mathcal P}_{\nu}^{\,\, \alpha}$. As far as the projection 
along $u_{\nu}$ is concerned we obtain 
\begin{equation}
u^{\alpha} \, \nabla_{\alpha} \bigl[ \rho \, f(\rho)] + \theta \,q(\rho) \,( \rho + p) =0, \qquad\qquad \theta = \nabla_{\mu} \, u^{\mu},
\label{CC1}
\end{equation}
where $\theta$ now denotes, as indicated, the total expansion. Equation (\ref{CC1}) implies that the covariant conservation of $T^{\mu\nu}$ reproduces exactly the condition derived from the energy-momentum tensor of Eq. (\ref{EQ2}) (i.e. $u_{\nu} \, \nabla_{\mu} {\mathcal t}^{\mu\nu}=0$)  provided
\begin{equation}
q(\rho) = \rho\,\, \partial_{\rho} f(\rho) + f(\rho), \qquad \partial_{\rho} \, f= \frac{\partial f}{\partial\rho}.
\label{CC2}
\end{equation}
Note that, in what follows, we shall often employ the notation $\partial_{\rho}$ to indicate
a derivation with respect to the energy density. 
The projection of $\nabla_{\mu} \, T^{\mu\nu} =0$ along ${\mathcal P}_{\nu}^{\,\,\alpha}$ gives instead the following condition 
\begin{equation}
(p+ \rho) q(\rho) u^{\mu}\, \nabla_{\mu} u^{\alpha} + {\mathcal P}_{\mu}^{\,\,\alpha} \partial^{\mu} \biggl[ \rho \, f(\rho) - (p + \rho) q(\rho)\biggr] =0.
\label{CC3}
\end{equation}
Therefore we have that the condition ${\mathcal P}_{\nu}^{\,\,\alpha} \nabla_{\mu} \, {\mathcal t}^{\mu\nu} =0$ is reproduced 
provided
\begin{equation}
\partial^{\mu}[ \rho f(\rho) - ( p + \rho) q(\rho)] = - q(\rho) \partial^{\mu} p.
\label{CC4}
\end{equation}
For a homogeneous background the only relevant condition is represented by Eq. (\ref{CC2}). 
In the case of perfect irrotational fluids the pressure and the energy density 
can be related via the barotropic index $w$ while the derivatives 
of the pressure and of the energy density define the square of the (total) sound speed $c_{s,\,t}^2$:
\begin{equation}
w= p/\rho, \qquad \qquad c_{s,\,t}^2 = \partial p/\partial\rho.
\label{CC4a}
\end{equation}
In terms of $c_{s,\,t}^2$ the fluctuations of the pressure can always 
be written as the sum of an adiabatic term supplemented by the 
non-adiabatic contribution:
\begin{equation}
\delta_{s} p= c_{s,\,t}^2 \delta_{s} \rho + \delta p_{nad},
\label{CC4b}
\end{equation}
where $\delta_{s} \rho$ and $\delta_{s} \, p$ denote the first-order (scalar) fluctuations 
of the energy density and of the pressure while $\delta p_{nad}$ accounts here for the non-adiabatic pressure fluctuations which are typically vanishing in the case of the adiabatic paradigm \cite{ADP1,ADP2}. The conditions (\ref{CC4a})--(\ref{CC4b}) become relevant for the  evolution of the curvature inhomogeneities, as we are going to see in section \ref{sec3}.

Before going to the case of scalar field matter we want to stress that the discussion of this section does not specifically 
assume any range for $w$. It is however relevant to mention that, in what follows, we shall assume 
explicitly or implicitly that $0\leq w \leq 1$. This choice implies, in particular, that none of the 
energy conditions are violated in the case of the original fluid appearing in Eq. (\ref{EQ2}).

\subsection{The case of scalar field matter}
The discussion of the perfect irrotational fluids can be  complemented by the case of scalar field matter 
where the role of the four-velocity is  played by the covariant gradients of the scalar field 
$\varphi$; in this subsection we are going to use the following stenographic notation:
\begin{equation}
u^{\mu} = \nabla^{\mu}\varphi/\sqrt{(\partial \varphi)^2}, \qquad \qquad (\partial \varphi)^2 = g^{\alpha\beta} \partial_{\alpha} \varphi\, 
\partial_{\beta} \varphi.
\label{DD1}
\end{equation}
The standard energy-momentum tensor associated with the scalar field $\varphi$ is the given by
\begin{equation}
{\mathcal t}_{\mu\nu}^{(\varphi)} = \partial_{\mu} \varphi \partial_{\nu} \varphi - g_{\mu\nu} \biggl[\frac{1}{2} g^{\alpha\beta} \partial_{\alpha} \varphi \partial_{\beta}\varphi - V(\varphi)\biggr],
\label{DD2}
\end{equation}
and if we now use Eq. (\ref{DD1}) inside Eq. (\ref{DD2}) we obtain 
the scalar field analog of  Eq. (\ref{EQ2})
\begin{equation}
t_{\mu\nu}^{(\varphi)} = - (\rho_{\varphi} + p_{\varphi}) {\mathcal P}_{\mu\nu} + \rho_{\varphi} \, g_{\mu\nu}.
\label{DD3}
\end{equation}
When projected along  ${\mathcal P}^{\mu\nu}$ and $u^{\mu}\, u^{\nu}$, Eqs. (\ref{DD2})--(\ref{DD3}) lead to the explicit forms of the pressure and of the energy density:
\begin{eqnarray}
p_{\varphi} &=& - \frac{1}{3} {\mathcal P}^{\mu\nu}\,\,{\mathcal t}_{\mu\nu} =  \frac{1}{2} \, g^{\alpha\beta} \partial_{\alpha} \varphi \partial_{\beta} \varphi  - V(\varphi),
\label{DD3aa}\\
\rho_{\varphi} &=& u^{\mu} \, u^{\nu} \, {\mathcal t}_{\mu\nu} =  \frac{1}{2} \, g^{\alpha\beta} \partial_{\alpha} \varphi \partial_{\beta} \varphi  + V(\varphi).
\label{DD3a}
\end{eqnarray}
If we now follow the logic outlined in the perfect fluid case,  the 
two smearing functions $f(\rho_{\varphi})$ and $q(\rho_{\varphi})$ can be introduced in the same manner so that, ultimately, the total energy-momentum tensor $T_{\mu\nu}^{(\varphi)} $
can be written as:
\begin{equation}
{\mathcal t}_{\mu\nu}^{(\varphi)} \qquad \Rightarrow\qquad  T_{\mu\nu}^{(\varphi)} = - (\rho_{\varphi} + p_{\varphi}) q(\rho_{\varphi}) {\mathcal P}_{\mu\nu} + 
\rho_{\varphi} \, f(\rho_{\varphi}) g_{\mu\nu}.
\label{DD4}
\end{equation}
Once more the covariant conservation of $T_{\mu\nu}^{(\varphi)}$ now implies the condition
\begin{equation}
\biggl[ f(\rho_{\varphi}) + \rho_{\varphi} \frac{\partial f}{\partial \rho_{\varphi}} \biggr] u^{\mu} \nabla \rho_{\varphi} + 
\theta \, q(\rho_{\varphi}) ( p_{\varphi} + \rho_{\varphi}) = 0,
\label{DD5}
\end{equation}
where, as in Eq. (\ref{CC1}), $\theta = \nabla_{\mu} \, u^{\mu}$ is the total expansion. In analogy with the case of a perfect relativistic fluid we can impose 
the same condition  and, in this case, Eq. (\ref{DD5}) implies the standard form of the Klein-Gordon equation:
\begin{equation}
f(\rho_{\varphi}) + \rho_{\varphi} \frac{\partial f}{\partial\rho_{\varphi}} = q(\rho_{\varphi}) \qquad \Rightarrow \qquad 
\sqrt{(\partial \varphi)^2} \biggl[ \nabla_{\mu} \nabla^{\mu} \varphi + \frac{\partial V}{\partial \varphi} \biggr] =0.
\label{DD6}
\end{equation}
Note that Eqs. (\ref{CC2}) and (\ref{DD6}) illustrate in fact the same condition written, however, for two physically different energy-momentum tensors. The analogies and the differences of the two cases have a direct counterpart in the evolution of the curvature inhomogeneities discussed in section \ref{sec3}. 

\subsection{Possible extensions}
Several extensions of the procedure suggested in the previous subsections can be envisaged both for isotropic and anisotropic backgrounds. We can note, for the sake of illustration, that the underlying fluid does not need to be inviscid and, in particular, the energy-momentum $T_{\mu\nu}$ can be  easily complemented by a viscous correction denoted hereunder by ${\mathcal T}_{\mu\nu}$. The total energy-momentum tensor will then be given 
by the usual inviscid contribution (containing the two smearing functions) 
supplemented by the viscous correction ${\mathcal T}_{\mu\nu}$:
\begin{equation}
T_{\mu\nu}^{(tot)} = - (p +\rho) \,q(\rho) \, {\mathcal P}_{\mu\nu} + \rho \, f(\rho) \, g_{\mu\nu} + {\mathcal T}_{\mu\nu}.
\label{EE1}
\end{equation}
As usual we may then parametrize the viscous contribution in terms of the first and second viscosities so that 
${\mathcal T}_{\mu\nu}$ can be written as:
\begin{equation}
{\mathcal T}_{\mu\nu} = 2\, \eta \,\sigma_{\mu\nu} + \xi\, (\nabla_{\alpha} \, u^{\alpha}) \, {\mathcal P}_{\mu\nu},
\label{EE2}
\end{equation}
where $\eta$ and $\xi$ are, by definition, the shear and bulk viscosity coefficients while $\sigma_{\mu\nu}$ denotes the standard form of the shear tensor:
\begin{equation}
\sigma_{\mu\nu} = \frac{1}{2} {\mathcal P}_{\mu}^{\,\,\alpha} {\mathcal P}_{\nu}^{\,\,\beta} {\mathcal B}_{\alpha\beta}, \qquad \qquad 
{\mathcal B}_{\alpha\beta} = \nabla_{\alpha} u_{\beta} + \nabla_{\beta} u_{\alpha} - \frac{2}{3} g_{\alpha\beta} \nabla_{\gamma} u^{\gamma}.
\label{EE3}
\end{equation}
It follows from Eqs. (\ref{EE2})-(\ref{EE3}) that the projection of ${\mathcal T}^{\mu\nu}$ along $u_{\mu}$ or $u_{\nu}$ is consistently vanishing, i.e. $u_{\mu} \, {\mathcal T}^{\mu\nu} = 
u_{\nu} \, {\mathcal T}^{\mu\nu} =0$.  We may finally introduce the diffusion current $j^{\mu}$ together with the dissipative contribution $\nu_{\mu}$:
\begin{equation}
j^{\mu} = n\, u^{\mu} + \nu^{\mu}, \qquad \qquad \nabla_{\mu} \, j^{\mu} =0.
\label{EE3a}
\end{equation}
In the Landau-Lifshitz approach \cite{LL,LL1} (which is the one followed here\footnote{For the sake of accuracy we mention that 
the Landau-Lifshitz frame is not the only one customarily adopted for viscous fluids; in the   Eckart approach (often employed in cosmology \cite{EE}) the $u_{\mu}$ does not 
describe the energy transport but rather the particle transport and it is fixed by requiring that 
$u_{\mu}\, j^{\mu} =0$ while $u_{\mu} {\mathcal T}^{\mu\nu} \neq 0$. }) the pure thermal 
conduction is related to an energy flux without particles  (since, as already noted,
$u_{\mu} \, {\mathcal T}^{\mu\nu} =0$). With these caveats, from the covariant conservation of the total energy-momentum tensor 
we obtain, as expected, 
\begin{equation}
\partial_{\rho} [ \rho\, f(\rho)] \, u^{\mu} \nabla_{\mu} \rho + \theta (p + \rho) q(\rho) = \xi \, \theta^2 + 2\, \eta\, \sigma^2, 
\label{EE4}
\end{equation}
where $\sigma^2 = \sigma_{\mu\nu} \, \sigma^{\mu\nu}$. Provided the condition (\ref{CC2}) holds, the viscous corrections lead to the same evolution of the entropy four-vector. Indeed, 
assuming for simplicity a vanishing chemical potential, the fundamental 
thermodynamic identity is ${\mathcal E} = T\,S - p\, V$ (where $V$ is the volume, $S$ 
is the entropy and ${\mathcal E}$ is the internal energy). Thus the entropy density is, by definition $\varsigma = (\rho + p )\, T$. If the fundamental thermodynamic identity 
is now combined with the first principle of thermodynamics (i.e. $d \, {\mathcal E} = T\, d s - p dV$) 
the gradients of the temperature and of the pressure 
are related as $ \varsigma \nabla_{\alpha}\, T = \nabla_{\alpha} p$;  thanks to the latter 
relation, the covariant divergence of the entropy four-vector $\varsigma^{\mu} = \varsigma \, u^{\mu}$ becomes
\begin{equation}
\nabla_{\mu} \varsigma^{\mu}  + \frac{1}{q\, T} \biggl\{ \frac{\partial}{\partial\rho} \bigl[ \rho f(\rho)\bigr] - q(\rho)\biggr\} = \frac{\xi}{q\, T} \theta^2 + 2 \frac{\eta}{q\, T} \sigma^2.
\label{EE5}
\end{equation}
From Eq. (\ref{EE5}) the entropy is conserved as long as
$\xi\to 0$, $\eta\to 0$ and provided the relation between $q(\rho)$ and $f(\rho)$ is 
fixed as in Eq. (\ref{CC2}). Moreover, in case Eq. (\ref{CC2}) would not be imposed 
the second principle of thermodynamics would not be generally valid. 
In the presence of a chemical 
potential $\mu$ the fundamental thermodynamic identity and the first 
principle of the thermodynamics are modified by the addition 
of  $\mu N$ and $\mu \, dN$ (where $N$ denotes the number of particles) 
at the right-hand side of the relations defining, respectively, ${\mathcal E}$ and $d {\mathcal E}$. As a consequence the gradient of the temperature and of the pressure are now related as $\varsigma \nabla_{\alpha} T - \nabla_{\alpha} p + n\, \nabla_{\alpha} \mu=0$ where $n$ now denotes the particle concentration. Using now 
Eq. (\ref{EE3a}) (and in particular the condition coming from $\nabla_{\mu}(n \, u^{\mu} + \nu^{\mu}) =0$) the generalization of Eq. (\ref{EE5}) becomes
\begin{equation}
\nabla_{\mu} \biggl[ \varsigma^{\mu}  - \frac{\mu}{T} \, u^{\mu}\biggr] 
+ \frac{\nu^{\alpha}}{T} +  \frac{1}{q\, T} \biggl\{ \frac{\partial}{\partial\rho} \bigl[ \rho f(\rho) \bigr] - q(\rho)\biggr\} = \frac{\xi}{q\, T} \theta^2 + 2 \frac{\eta}{q\, T} \sigma^2.
\label{EE6}
\end{equation}
When Eq. (\ref{CC2}) is enforced we obtain 
the standard result which however contains an explicit dependence on $q(\rho)$ 
in the viscous contribution. This means, as expected, that Eq. (\ref{EE6}) coincides with the standard 
result provided $q(\rho) \to 1$.

\renewcommand{\theequation}{3.\arabic{equation}}
\setcounter{equation}{0}
\section{Evolution of the curvature inhomogeneities}
\label{sec3}
 In the case of a perfect irrotational fluid the system of the Einstein's equation supplemented by the covariant conservation can be written in a Friedmann-Robertson-Walker metric:
\begin{eqnarray}
 && {\mathcal H}^2 = a^2 \frac{\ell_{P}^2}{3} \rho \, f(\rho) - \kappa,
 \label{CC6}\\
&& {\mathcal H}^2 - {\mathcal H}^{\prime} = a^2 \frac{\ell_{P}^2}{2} (p + \rho) q(\rho) - \kappa,
\label{CC7}\\
&& \rho^{\prime} + 3 {\mathcal H} \frac{q(\rho)}{\partial_{\rho} [ \rho f(\rho)]} (\rho + p) =0,
\label{CC8}
\end{eqnarray}
where $\kappa$ denotes the spatial (intrinsic) curvature; the prime accounts for the derivation with respect to the conformal time coordinate $\tau$ and ${\mathcal H} = a^{\prime}/a$. As long as the condition (\ref{CC2}) is  imposed, Eq. (\ref{CC8}) reduces to $\rho^{\prime} + 3 {\mathcal H} (\rho +p)=0$ and this means 
that the covariant conservation keeps its original form even in the presence 
of smearing functions $f(\rho)$ and $q(\rho)$. Although the condition (\ref{CC2}) eliminates 
the smearing functions from Eq. (\ref{CC8}), the Hubble rate and its derivative 
are anyway affected by $f(\rho)$ and $q(\rho)$. The same observations hold, with some technical 
differences, for the scalar field matter where, for in a homogenous background Eqs.  (\ref{DD3aa})--(\ref{DD3a})
read 
\begin{equation}
p_{\varphi} = \frac{\varphi^{\prime\, 2}}{2 a^2 } - V(\varphi), \qquad\qquad \rho_{\varphi} = \frac{\varphi^{\prime\, 2}}{2 a^2 } + V(\varphi).
\label{CC9}
\end{equation}
Thanks to Eq. (\ref{CC9}) the scalar field version of Eqs. (\ref{CC6})--(\ref{CC7}) now becomes
\begin{eqnarray}
 && 3 {\mathcal H}^2 = a^2 \ell_{P}^2\rho_{\varphi} \, f(\rho_{\varphi}) - 3\,\kappa,
 \label{CC6a}\\
&& 2( {\mathcal H}^2 - {\mathcal H}^{\prime} )= \ell_{P}^2 \, q(\rho_{\varphi})\, \varphi^{\prime\, 2}  - 2\, \kappa.
\label{CC7a}
\end{eqnarray}
If the condition of Eq. (\ref{DD6}) is enforced the covariant conservation 
and the Klein-Gordon equations are equivalent and can be written, in the present case, as:
\begin{equation}
\rho_{\varphi}^{\prime} + 3 {\mathcal H} (\rho_{\varphi} + p_{\varphi}) =0\qquad \Rightarrow \qquad \varphi^{\prime\prime} + 2 {\mathcal H} \varphi^{\prime} + a^2 \, V_{,\, \varphi} =0, \qquad\qquad V_{,\, \varphi} = \frac{\partial V}{\partial \varphi}.
\label{CC10}
\end{equation}
In the class of scenarios discussed here the intrinsic (spatial) curvature is suppressed 
 prior to bouncing regime; this means in practice that $\kappa/{\mathcal H}^2 \ll 1$ around the maximal curvature scale. 
 
 \subsection{Practical gauge choices}
Since the curvature inhomogeneities are invariant under infinitesimal coordinate transformations, their evolution can be notoriously deduced in any specific gauge (see, for instance, \cite{WAD1,WAD2,WAD3}). In the present context a practical gauge choice is given by: 
\begin{equation}
\delta_{s} g_{00} = 2 \, a^2 \, \phi, \qquad \delta_{s} g_{i\, j} = 0, \qquad \delta_{s} g_{0i} = - a^2 \partial_{i} B,
\label{FF1}
\end{equation}
where, as in Eq. (\ref{CC4b}),  $\delta_{s}$ denotes the first-order (scalar) fluctuation of the corresponding quantity.
The condition $\delta_{s} g_{ij} =0$ completely fixes the gauge freedom; indeed, $\delta_{s} g_{i\, j}$  can be 
written, in four-dimensions, as $\delta_{s} g_{i\, j} = 2\, a^2 (\psi \delta_{i\, j} - \partial_{i} \, \partial_{j} E)$. For infinitesimal coordinate shifts transformations of the type $\tau \to \overline{\tau} = \tau + \epsilon^{0}$ and $x^{i} \to \overline{x}^{i} = x^{i} + \epsilon^{i}$ the metric fluctuations transform as:
\begin{equation}
\psi \to \overline{\psi} = \psi + {\mathcal H} \epsilon_{0}, \qquad E \to \overline{E} = E - \epsilon, 
\label{FF2}
\end{equation}
where $\epsilon_{\mu} = a^2(\epsilon_{0}, - \partial_{i} \epsilon)$. Equation (\ref{FF2}) implies 
that if we start from a gauge where $\delta_{s} g_{i\, j} \neq 0$ (i.e. $\psi \neq 0$ and $E \neq 0$), we can always impose the coordinate system of Eq. (\ref{FF1}) by requiring $\epsilon = E$ and ${\mathcal H}\epsilon_{0} = - \psi $. In the gauge (\ref{FF1}) the curvature inhomogeneities on comoving orthogonal hypersurfaces (see also Eqs. (\ref{APPA7})--(\ref{APPA8}) of appendix \ref{APPA}) are then defined as
\begin{equation}
{\mathcal R} = - \frac{{\mathcal H}^2}{{\mathcal H}^2 - {\mathcal H}^{\prime}} \phi = \frac{H^2}{\dot{H}} \phi,
\label{FF3}
\end{equation}
where the second equality follows from the standard connection between 
conformal and cosmic time parametrizations. In what follows
the evolution of ${\mathcal R}$ will be deduced in terms of the two smearing functions $f(\rho)$ and $q(\rho)$. 

\subsection{Inhomogeneities of the fluid sources}
In the case of the perfect irrotational fluids the scalar fluctuation of the 
energy-momentum tensor of Eq. (\ref{EQ3}) can be expressed, in a covariant form, as:
\begin{eqnarray}
\delta_{s} \, T_{\mu}^{\,\,\nu} &=& - \bigl[ (\delta_{s} p + \delta_{s}\, \rho) \, q(\rho) + (p + \rho)\, \partial_{\rho} q \,\,\delta_{s} \rho\bigr] {\mathcal P}_{\mu}^{\,\,\nu} 
\nonumber\\
&+& \partial_{\rho} \bigl[ \rho\, f(\rho)\bigr] \delta_{s}\rho \,\, \delta_{\mu}^{\,\,\nu} - (p +\rho) \,q(\rho) \, \delta_{s}  {\mathcal P}_{\mu}^{\,\, \nu}.
\label{FF4}
\end{eqnarray}
Although we have that  $\delta_{s} {\mathcal P}_{\mu}^{\,\,\, \nu}= \overline{u}_{\mu} \delta_{s} \, u^{\nu} + \delta_{s} u_{\mu} \overline{u}^{\nu}$, in the gauge\footnote{Note that, by definition, $\overline{g}_{\mu\nu} \, \overline{u}^{\mu}\, \overline{u}^{\nu} =1$; this means that $\overline{u}_{0} = a$ while $\overline{u}_{i} =0$; in the gauge (\ref{FF1}) $\delta_{s} u_{0} = a \,\phi$  and $ \delta u^{0} = - \phi/a$; this is why, ultimately, $\delta_{s} {\mathcal P}_{0}^{\,\,\, 0}= \delta_{s} {\mathcal P}_{i}^{\,\,\, j} \,= \,0$.} of Eq. (\ref{FF1}) $\delta_{s} {\mathcal P}_{0}^{\,\,\, 0}= \delta_{s} {\mathcal P}_{i}^{\,\,\, j}=0 $ whereas $\delta_{s} {\mathcal P}_{i}^{\,\,\, 0} = (p+ \rho) \, q(\rho) u_{0} \delta\, u_{i}$ 
and similarly for  $\delta_{s} {\mathcal P}_{0}^{\,\,\, i}$. This 
observation implies that the various components of $\delta_{s} \, T_{\mu}^{\,\,\,\nu}$  are:
\begin{eqnarray}
\delta_{s} \, T_{0}^{\,\,0} &=& \biggl[ f(\rho) + \rho \, \frac{\partial f}{\partial \rho} \biggr] \delta_{s} \rho, 
\label{FF5}\\
\delta_{s} \, T_{i}^{\,\,j} &=& - \bigl[ ( \delta_{s} p + \delta_{s} \rho) q(\rho) + (p + \rho)\, \partial_{\rho} q \, \delta_{s} \rho \bigr] \,\delta_{i}^{\,\, j}
\nonumber\\
&+& \biggl[ f(\rho) + \rho \frac{\partial f}{\partial \rho} \biggr] \delta_{s} \rho \,\delta_{i}^{\,\,j} + \Pi_{i}^{\,\,j},
\label{FF6}\\
\delta_{s} T_{i}^{\,\,0} &=& (p + \rho) \, q(\rho) \, u^{0} \,\delta u_{i}.
\label{FF7}
\end{eqnarray}
In Eq. (\ref{FF6}) we added, for the sake of completeness, the (traceless) anisotropic stress $\Pi_{i}^{\,\, j}$. 
Inserting Eq. (\ref{CC4b}) into Eq. (\ref{FF6}) we obtain the following form of the spatial fluctuations of the energy-momentum 
tensor where the non-adiabatic pressure fluctuations $\delta p_{nad}$ explicitly appear:
\begin{equation}
\delta_{s} \, T_{i}^{\,\,j} = - \biggl\{ ( c_{s,\,t}^2 + 1) \, q(\rho) + (p + \rho)\, \frac{\partial q}{\partial\rho} -\frac{\partial }{\partial \rho}\bigl[ \rho\, f(\rho)\bigr]\biggr] \biggr\}\, \delta_{s} \rho + q(\rho) \, \delta p_{nad} + \Pi_{i}^{\,\,j}.
\label{FF8}
\end{equation}
The simplest way to derive the evolution of the curvature inhomogeneities of Eq. (\ref{FF3}) is to start by partially 
decoupling the equations that involve $\phi$. Along this perspective the $(0, 0)$ component 
of perturbed Einstein's equations follows from Eq. (\ref{FF5}) and from the corresponding 
fluctuation of the Einstein's tensor reported in Eq. (\ref{APPA1}):
\begin{equation}
- {\mathcal H} \nabla^2 B - 3 {\mathcal H}^2 \phi = \frac{\ell_{P}^2\, a^2}{2} \biggl[ f + \rho \, \frac{\partial f}{\partial \rho} \biggr] \,\delta_{s} \rho.
\label{FF9}
\end{equation}
The spatial component of the perturbed Einstein's equations imply a second independent relation; in particular from Eqs. (\ref{FF6}) and (\ref{APPA2}) we get:
\begin{eqnarray}
&&\frac{1}{a^2} \biggl\{ \bigg[ - 
2 ({\mathcal H}^2 + 2 {\mathcal H}') \phi - 2 {\mathcal H} \phi' \biggr]
- \nabla^2 (  B^{\prime} + 2 {\mathcal H} B + \phi)\biggr\} \delta_{i}^{\,\,j}
\nonumber\\
&&+ \frac{1}{a^2}\partial_{i}\partial^{j} \biggl[  B^{\prime} + 2 {\mathcal H} B + \phi \biggr] = \ell_{P}^2 \biggl\{
- \bigl[ ( \delta_{s} p + \delta_{s} \rho) q(\rho) + (p + \rho)\, \partial_{\rho} q \, \delta_{s} \rho \bigr] \,\delta_{i}^{\,\, j}
\nonumber\\ 
&&+ \biggl[ f(\rho) + \rho \frac{\partial f}{\partial \rho} \biggr] \delta_{s} \rho \,\delta_{i}^{\,\,j} + \Pi_{i}^{\,\,j}\biggr\}.
\label{FF9a}
\end{eqnarray}
If we now trace of both sides of Eq. (\ref{FF9a}) we can immediately deduce
\begin{eqnarray}
({\mathcal H}^2 + 2 {\mathcal H}^{\prime}) \phi +  {\mathcal H} \phi^{\prime} + \frac{1}{3}\nabla^2 {\mathcal Q} = \frac{\ell_{P}^2 \, a^2}{2}\biggl\{ (\delta_{s} p + \delta_{s} \rho)q 
+ \biggl[(p +\rho) \partial_{\rho} q - \partial_{\rho}(\rho f) \biggr] \delta_{s} \rho\, \biggr\},
\label{FF10}
\end{eqnarray}
where, for the sake of conciseness, we introduced the combination ${\mathcal Q} =(\phi + B^{\prime} + 2 {\mathcal H} B)$; for the same reason we also wrote $f \equiv f(\rho)$ and $q\equiv q(\rho)$ and suppressed the arguments
since it is clear that the smearing functions only depend upon the energy density. 
If we then subtract Eq. (\ref{FF10}) from Eq. (\ref{FF9a}) we obtain that the combination appearing inside the Laplacian at the right hand side of Eq. (\ref{FF10}) differs from zero only in the presence of an anisotropic stress; in other words from the traceless  contributions to the spatial components of the perturbed Einstein's equations we can easily obtain 
\begin{equation}
\partial_{i}\partial^{j} {\mathcal Q} - \frac{1}{3} \nabla^2\, {\mathcal Q} \, \delta_{i}^{\,\,j}= \ell_{P}^2 \, a^2 \, \Pi_{i}^{\,\,j}.
\label{FF11}
\end{equation}
By now applying $\partial_{j}\, \partial^{i}$ to both sides of Eq. (\ref{FF11}) we can finally deduce that 
\begin{equation}
\nabla^4 (B^{\prime} + 2 {\mathcal H} B + \phi) = \frac{3}{2} \ell_{P}^2 a^2  \nabla^2 \Pi ,
\label{FF12}
\end{equation}
where  $\Pi$ is implicitly defined as $\nabla^2 \Pi = \partial_{j} \partial^{i} \,\Pi_{i}^{\,\,j}$. It is useful to recall that 
 in the gauge (\ref{FF1}) the Bardeen potentials \cite{BBR1} 
read $\Psi = - {\mathcal H} \, B$ and $ \Phi = \phi + {\mathcal H} B + B^{\prime}$. Since the anisotropic stress is gauge-invariant 
because of the Stewart-Walker lemma \cite{SWL}, we also have that the gauge-invariant expression of Eq. 
(\ref{FF12}) is, as expected, $\nabla^4 (\Phi - \Psi) = 12 \pi G a^2 \nabla^2 \Pi$. Similarly the expression of the curvature inhomogeneities of Eq. (\ref{FF3}) can be easily expressed in terms of the 
Bardeen potentials: 
\begin{equation} 
{\mathcal R} = - \Psi - \frac{{\mathcal H}( {\mathcal H} \Phi + \Psi^{\prime})}{{\mathcal H}^2 - {\mathcal H}^{\prime}}.
\label{FF13}
\end{equation}
If we now recall, as in the case of Eq. (\ref{FF12}) the connection between the gauge-invariant 
potentials (i.e. $\Phi$ and $\Psi$) and the gauge-dependent variables (i.e. $\phi$ and $B$) 
 the expression already mentioned in Eq. (\ref{FF3}) can be easily reobtained from Eq. (\ref{FF13}).

\subsection{Evolution of the curvature inhomogeneities}
Equations (\ref{FF9}) and (\ref{FF10}) can be  decoupled in terms of $\phi$ and $B$: after eliminating the fluid sources between the two equations we may trade $\phi$ for ${\mathcal R}$ according to Eq. (\ref{FF3}). The final result of this procedure becomes\footnote{In Eq. (\ref{FF14}) ${\mathcal S}_{{\mathcal R}}$ depends on the non-adiabatic pressure 
fluctuations of the fluid and also on the total anisotropic stress $\Pi$. It is actually simpler to incorporate all the sources in a single notations. During an inflationary or bouncing stage the anisotropic stress can be neglected especially for typical wavelengths 
larger than the Hubble radius. Furthermore, as suggested earlier on in the introduction, we shall 
be mainly interested in the situations that are, broadly speaking, compatible with the adiabatic paradigm \cite{ADP1,ADP2}.}:
\begin{eqnarray}
{\mathcal R}^{\prime} &=& \frac{2 \, {\mathcal H}^2\, c_{eff}^2 \,\nabla^2 B }{ \ell_{P}^2 a^2 (\rho + p) \,q(\rho) } + {\mathcal S}_{\mathcal R},
\label{FF14}\\
{\mathcal S}_{{\mathcal R}} &=& - \frac{{\mathcal H}}{ q(\rho) (p + \rho)}\biggl[ q(\rho)\, \delta p_{nad} - \Pi\biggr].
\label{FF17}
\end{eqnarray}
where $c_{eff}^2$ is an effective sound speed that is related both to $c_{s,\,t}^2$ and to the smearing functions:
\begin{equation}
c_{eff}^2(\rho,\,c_{s,\,t}^2) = \frac{q(\rho) \, c_{s,\,t}^2}{\partial_{\rho}[ \rho\, f(\rho)]} + \biggl\{ \frac{q(\rho)}{\partial_{\rho}[\rho \, f(\rho)]} - 1\biggr\} + \biggl(1 + \frac{p}{\rho}\biggr) \frac{ \rho \,\,\partial_{\rho} q}{\partial_{\rho} [ \rho f(\rho)]}.
\label{FF15}
\end{equation}
As stressed before, when Eq. (\ref{CC2}) is enforced $ q(\rho) = \partial_{\rho}[ \rho f(\rho)]$ and the covariant conservation 
remains unmodified in comparison with the case $q(\rho) \to 1$ and $f(\rho) \to 1$; with this choice 
 Eq. (\ref{FF15}) simplifies and  becomes
\begin{equation}
c_{eff}^2 = c_{s,\,t}^2 + ( 1 + w)  \, \biggl(\frac{\rho}{q}\biggr) \biggl(\frac{\partial q}{\partial \rho}\biggr),
\label{FF16}
\end{equation}
where barotropic index $w$ has been introduced. In the limit $q(\rho) \to 1$ 
Eq. (\ref{FF16}) implies, as expected, that $c_{eff}^2 \to c_{s,\,t}^2$; this 
is consistent with the standard result originally obtained in Refs. \cite{BBR2,BBR2b,BBR2c}.
Since one of the two Bardeen potentials is related to $B$ as $\Psi = - {\mathcal H} B$, Eq. (\ref{FF14}) can also be written as
\begin{equation}
{\mathcal R}^{\prime} = - \frac{ {\mathcal H}\, c_{eff}^2 \,\nabla^2 \Psi }{ 4 \pi G a^2 (\rho + p) \,q(\rho) } + {\mathcal S}_{{\mathcal R}},
\label{FF18}
\end{equation}
where we also traded $\ell_{P}^2$ for $8 \pi G$. If the Laplacian 
at the right-hand side of Eq. (\ref{FF18}) are neglected the resulting equation 
accounts for the approximate evolution of the curvature inhomogeneities  the large-scale limit.
This is not what we are going to do since, as we shall see, it is easier to discuss the general 
situation after by using the decoupled evolution for ${\mathcal R}$. For this purpose we now introduce 
a background variable conventionally denoted by $z$:
\begin{equation}
z^2 = \frac{a^4 \, q(\rho) ( p + \rho)}{{\mathcal H}^2 \, c_{eff}^2(\rho)}.
\label{FF19}
\end{equation}
In terms of $z^2$, Eq. (\ref{FF14}) can be rephrased in the following manner:
\begin{equation}
{\mathcal R}^{\prime} - {\mathcal S}_{{\mathcal R}}= \frac{a^2}{4 \pi G\, z^2} \nabla^2 B. 
\label{FF20}
\end{equation}
Now the strategy is very simple and can be summarized, in short, as follows. After taking the derivative of both sides of Eq. (\ref{FF20}), there will appear terms proportional to $\nabla^2B$ and to $\nabla^2 B^{\prime}$;
the new contributions can be eliminated in favour of ${\mathcal R}$ and ${\mathcal R}^{\prime}$ 
by using again Eqs. (\ref{FF12}) and (\ref{FF20}). If we now apply the first step of the strategy we get:
\begin{equation}
{\mathcal R}^{\prime\prime} - {\mathcal S}_{{\mathcal R}}^{\prime} = \biggl( 2 \frac{z^{\prime}}{z} - 2 {\mathcal H}\biggr) ({\mathcal R}^{\prime} - 
{\mathcal S}_{{\mathcal R}}) + \frac{1}{ 4 \pi G} \biggl(\frac{a^2}{z^2}\biggr) \nabla^2 B^{\prime}. 
\label{FF21}
\end{equation}
We can now eliminate the term containing $\nabla^2 B^{\prime}$ via Eq. (\ref{FF12}) but three new terms will appear: one proportional to $\nabla^2 B$ (which can be eliminated with Eq. (\ref{FF20})), one proportional to $\phi$ (and hence to ${\mathcal R}$, as dictated by Eq. (\ref{FF3})) and the last one containing the total anisotropic stress. After some 
simple algebraic simplifications the final form of Eq. (\ref{FF21}) becomes:
\begin{equation}
{\mathcal R}^{\prime\prime} + 2 \frac{z^{\prime}}{z} {\mathcal R}^{\prime} - c_{eff}^2 \nabla^2 {\mathcal R} 
= {\mathcal S}_{{\mathcal R}}^{\prime} + 2 \frac{z^{\prime}}{z} {\mathcal R}^{\prime} + 
\frac{3 a^4}{z^2} \Pi.
\label{FF22}
\end{equation}
Equation (\ref{FF22}) can then be analyzed in different situations but when the anisotropic stress 
and the non-adiabatic pressure fluctuations are both vanishing
(i.e. $\Pi \to 0$ and $\delta p_{nad} \to 0$) we have that ${\mathcal S}_{{\mathcal R}}\to 0$ and 
all the terms at the right-hand side of Eq. (\ref{FF22}) disappear:
\begin{equation}
{\mathcal R}^{\prime\prime} + 2 \frac{z^{\prime}}{z} {\mathcal R}^{\prime} - 
c_{eff}^2 \nabla^2 {\mathcal R} = 0.
\label{FF23}
\end{equation}
Equation (\ref{FF23}) is relevant at early times and during a bouncing stage; furthermore it is also the appropriate one in the absence of entropic fluctuations. This is actually the case broadly compatible with the adiabatic paradigm \cite{ADP1,ADP2}, as already mentioned earlier on.
When $q \to 1$ and $f\to 1$ Eqs. (\ref{FF19}) and (\ref{FF23}) coincide with the normal modes of  a gravitating, irrotational and relativistic fluid derived in Ref.  \cite{BBR2} (see also \cite{BBR2b,BBR2c}) even prior to  formulation of the  inflationary paradigm. As already stressed above, the variable ${\mathcal R}$ is  gauge-invariant an it coincide with the curvature perturbation on comoving orthogonal hypersurface (see, for instance,  \cite{BBR3,BBR4}). Equation (\ref{FF23}) will now be analyzed in two general situations corresponding, respectively, to the bounces of the scale factor and to the curvature bounces.

\subsubsection{Bounces of the scale factor}
As already stressed before in the introduction we shall generally consider two complementary physical situations. When the extrinsic (Hubble) curvature locally vanishes we are in the situation of the bounces of the scale factor where ${\mathcal H} \to 0$; in this case also $f(\rho)$ must vanish
and therefore a particular form of $f(\rho)$ pins down a specific class of solutions. In this section we shall only assume that $f(\rho)$ does vanish 
while in section \ref{sec4} a number of explicit solutions will be considered. The second complementary physical possibility is the one of the curvature bounces where the Hubble rate does not vanish but  its time derivative goes in fact to zero. Both cases will now be analyzed in succession. 

Let us then start with the bounces of the scale factor and, for this purpose, we can note that a
 necessary condition for the bounce of the scale factor to occur is that $f(\rho)$ vanishes 
 for some value of $\rho$ (be it $\rho_{1}$) so that,  as implied by Eq. (\ref{CC6}),  also ${\mathcal H}^2$ 
 will vanish in the same limit:
\begin{equation}
\lim_{\rho \to \rho_{1}} f(\rho) \to 0\qquad \Rightarrow\qquad \lim_{\tau\to \tau_{1}} {\mathcal H}^2 \to 0.
\label{GG1}
\end{equation}
While Eq. (\ref{GG1}) can be realized in different ways (and more explicit examples will be presented 
in section \ref{sec4}) we are now going to argue that in spite of the specific background dependence the zeros of $f(\rho)$ always imply that $c_{eff}^2 < 0$ for a finite range of the energy density.  For this 
purpose we go back to Eq. (\ref{FF16}) and consider the following parametrization\footnote{We could 
also parametrize $f(\rho)$ as $f(\rho) = (b - \rho/\rho_{1})^{\gamma}$ where  $b$ is a numerical factor different 
from $1$. It is always possible, in this situation, to rescale $b$ outside of the bracket. The typical energy $\rho_{1}$ 
will then be rescaled as $\overline{\rho}_{1} = b \rho_{1}$ and the parametrization becomes again 
of the type of the one given in Eq. (\ref{GG2}).}
of $f(\rho)$ and $q(\rho)$:
\begin{equation}
f(\rho) = \biggl(1 - \frac{\rho}{\rho_{1}}\biggr)^{\gamma}, \qquad q(\rho)= \biggl(1 - \frac{\rho}{\rho_{1}}\biggr)^{\gamma -1} \biggl[ 1 - (\gamma +1)\biggl(\frac{\rho}{\rho_{1}}\biggr)\biggr],
\label{GG2}
\end{equation}
where $0 < (\rho/\rho_{1}) \leq 1$. The value of $q(\rho)$  follows from the 
condition of Eq. (\ref{CC2}) and if Eq. (\ref{GG2}) is inserted into Eq. (\ref{FF16}) the explicit 
value of $c_{eff}^2$ becomes:
\begin{equation}
c_{eff}^2(r) = c_{s,\,t}^2 - (1+w) \frac{r\, [2 \gamma - r \,\gamma(\gamma+1)]}{(1-r) [ 1 - (\gamma+1)\,r]}
,\qquad r = \rho/\rho_{1}.
\label{GG3}
\end{equation}
Equation (\ref{GG3}) demands that, for different values of $w$ and $\gamma$, the square of the effective sound speed is negative (i.e. $c_{eff}^2 <0$) for $r = \rho/\rho_{1}$ 
ranging between $0$ and $1$. The appearance of an instability is then necessary since, according to Eq. (\ref{FF23}), the wavelengths shorter than the Hubble radius are exponentially amplified and invalidate the perturbative expansion. 
\begin{figure}[!ht]
\centering
\includegraphics[height=6.7cm]{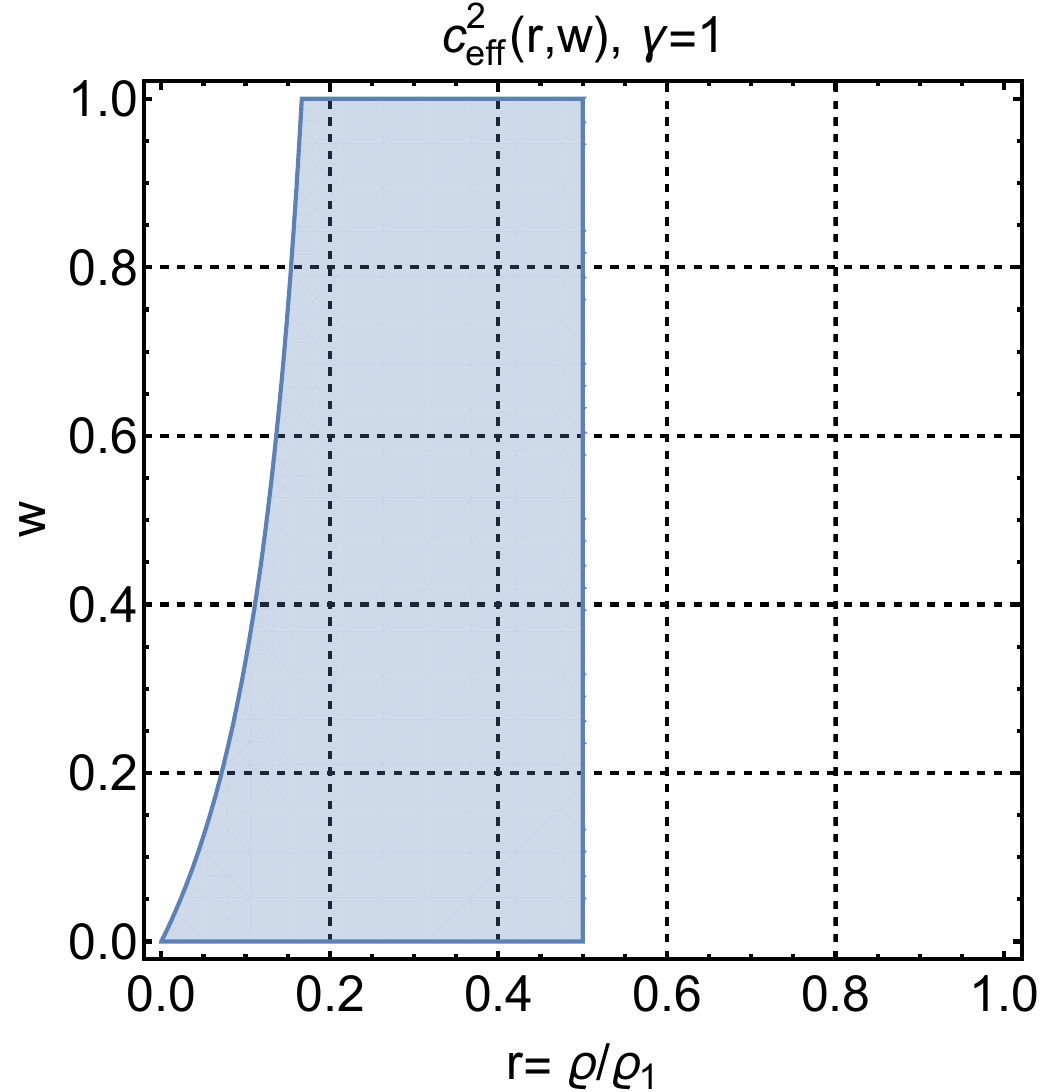}
\includegraphics[height=6.7cm]{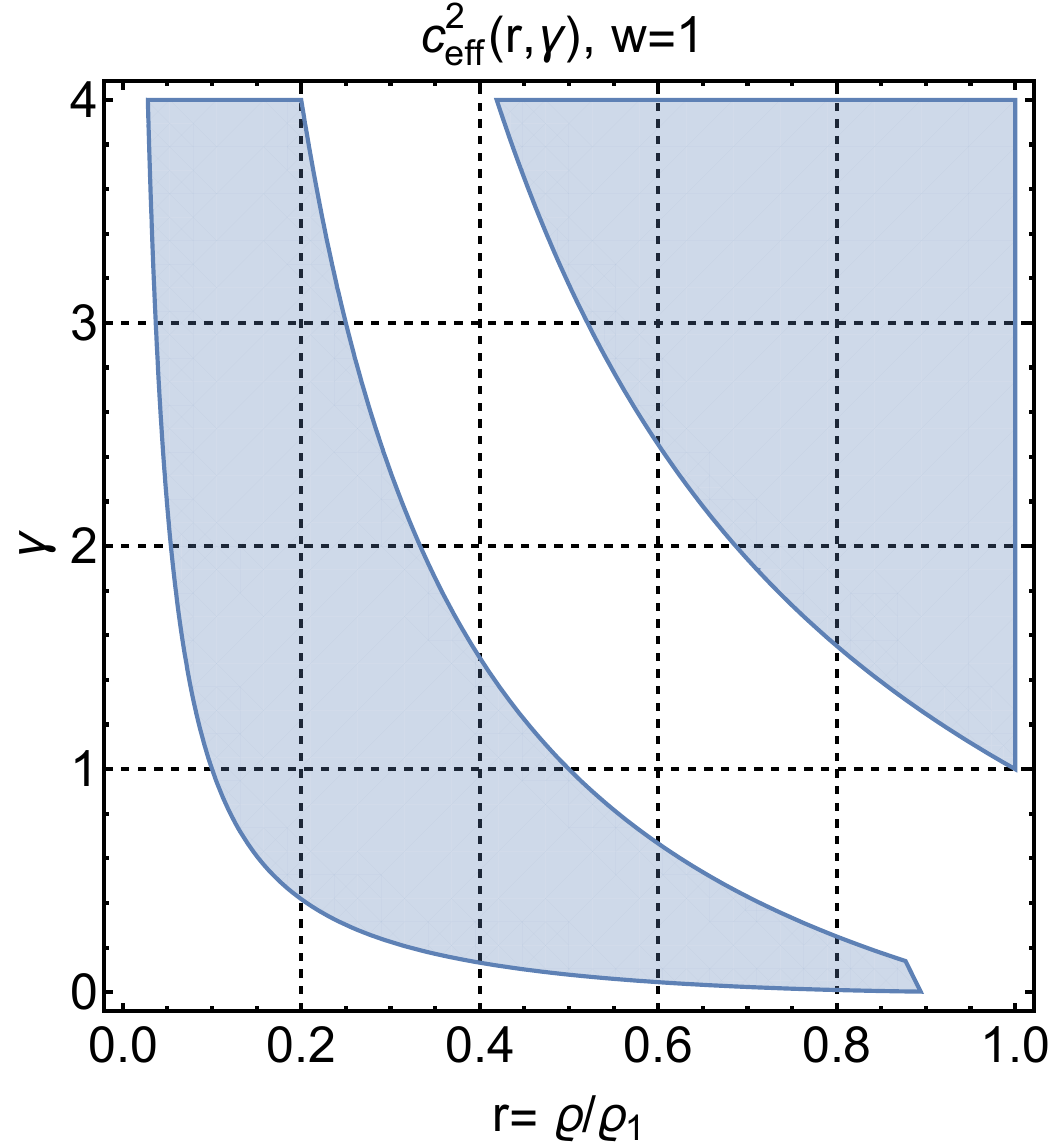}
\caption[a]{In both plots the shaded area illustrate the regions 
where $c_{eff}^2<0$ for different values of $w$ (the barotropic index) and $\gamma$ (the order of the zero of $f(\rho)$). Since $c_{eff}^2$ appears in front of the Laplacian of Eq. (\ref{FF23}), the shaded areas correspond in fact to an unstable evolution occurring in the small-scale limit. We stress that the preferred values of $w$ will be such that $0\leq w\leq 1$ to avoid violations of the energy conditions in the original fluid of Eq. (\ref{EQ2}).}
\label{FIGURE1}      
\end{figure}
Equation (\ref{GG3}) is illustrated in Fig. \ref{FIGURE1} where the shaded areas in both plots corresponds to the regions where $c_{eff}^2$ gets negative. It is actually well known that bouncing models may experience this kind of instability implying that 
 the scalar modes of the geometry inherit an imaginary sound speed \cite{BB2}. Sometimes the instability is dubbed gradient instability since, as in the 
case of Eq. (\ref{FF23}), the imaginary sound speed appears in front of the Laplacian.
Depending on the specific scenario these instabilities may be generic or not \cite{GI1} (see also 
\cite{GI2,GI3,GI4}). Equation (\ref{GG3}) and Fig. \ref{FIGURE1} suggest that, in the present context, the gradient instability is generic: as the energy density decreases from its maximal value in the bouncing region there will always be a region where $c_{eff}^2 <0$ unless the variation of $r$ is artificially restricted. 

We just showed that whenever $c_{eff}^2(\tau)< 0$ there are instabilities associated with the dominance of the spatial gradients and this happens for typical wavelengths shorter than the rate of variation of the geometry. Let us now investigate the opposite limit Eq. (\ref{FF23}) and show that  it also implies the existence of singularities. To clarify this point the evolution in Fourier space can be specifically analyzed and Eq. (\ref{FF23}) can then be rewritten as:
\begin{equation}
\partial_{\tau} \biggl[ z^2(\tau)\, \partial_{\tau} \, {\mathcal R}_{k} \biggr] + k^2 \, c_{eff}^2(\tau) z^2(\tau)\, {\mathcal R}_{k} =0.
\label{GG3a}
\end{equation}
By integrating once Eq. (\ref{GG3a}) we formally obtain an expression for the first 
derivative of the curvature inhomogeneities:
\begin{equation}
{\mathcal R}_{k}^{\prime}(\tau) = {\mathcal R}_{k}^{\prime}(\tau_{ex}) \frac{z_{ex}^2}{z^2(\tau)} - \frac{k^2}{z^2(\tau)} \int_{\tau_{ex}}^{\tau} \frac{a^4\, (\rho + p)\, q(\rho)}{{\mathcal H}^2}\biggl|_{\tau_{1}} {\mathcal R}_{k}(\tau_{1}) \,\, d \, \tau_{1},
\label{GG3b}
\end{equation}
where, by definition, $\tau > \tau_{ex} = {\mathcal O}(1/k)$; in the same time range  
Eq. (\ref{GG3b}) can be further integrated with respect to $\tau$ and after some trivial algebraic 
rearrangements the following result is finally obtained:
\begin{eqnarray}
{\mathcal R}_{k}(\tau) &=& {\mathcal R}_{k}(\tau_{ex}) + {\mathcal R}_{k}^{\prime}(\tau_{ex}) \int_{\tau_{ex}}^{\tau} \frac{z_{ex}^2}{z^2(\tau_{1})} \,\, d \tau_{1}
\nonumber\\
&-&  k^2 \, \int_{\tau_{ex}}^{\tau} \frac{d \tau_{1}}{z^2(\tau_{1})}\, \int_{\tau_{ex}}^{\tau_{1}} \frac{a^4\, (p+ \rho) q(\rho)}{{\mathcal H}^2}\biggl|_{\tau_{2}} \, {\mathcal R}_{k}(\tau_{2}) \, d\tau_{2}.
\label{GG3c}
\end{eqnarray}
Equation (\ref{GG3c}) is in fact an integral equation that can be iteratively solved so that the leading order result follows by replacing  ${\mathcal R}_{k}(\tau_{2})$ with  ${\mathcal R}_{k}(\tau_{ex})$:\begin{eqnarray}
{\mathcal R}_{k}(\tau) &=& {\mathcal R}_{k}(\tau_{ex}) + {\mathcal R}_{k}^{\prime}(\tau_{ex}) \int_{\tau_{ex}}^{\tau} \frac{z_{ex}^2}{z^2(\tau_{1})} \,\, d \tau_{1}
\nonumber\\
&-& 3 \, \overline{M}_{P}^2 (1+ w) \, k^2 \,  {\mathcal R}_{k}(\tau_{ex})  \int_{\tau_{ex}}^{\tau} \frac{d \tau_{1}}{z^2(\tau_{1})}\, \int_{\tau_{ex}}^{\tau_{1}} \frac{a^4(\tau_{2})\, q(\tau_{2})}{f(\tau_{2})}\,\, d\tau_{2}.
\label{GG3d}
\end{eqnarray}
In the large-scale limit  the third term of Eqs. (\ref{GG3c})--(\ref{GG3d}) will then negligible 
and hopefully finite\footnote{The third term in Eqs. (\ref{GG3c})--(\ref{GG3d}) may in fact diverge because of the gradient instability; however, to be conservative, we shall be assuming that it is finite and this may happen if the background solutions are artificially tuned within one of the stability regions appearing in Fig. \ref{FIGURE1}.}.  Even granting the absence of the gradient instabilities, 
the first derivative of ${\mathcal R}_{k}(\tau)$ is potentially divergent since it goes as $1/z^2(\tau)$.
We may now drop all the finite terms appearing in $1/z^2(\tau)$ and focus on the 
divergent contributions; in this perspective, for instance, $a^4(\tau)$ 
constitutes a regular contribution since we are here supposing that there exist non-singular bouncing solutions with the appropriate 
boundary conditions. Although explicit examples along this direction 
are in fact collected in section \ref{sec4}, what matters here are not the specific features of the background. 
With this simplification we then 
have that the divergences of ${\mathcal R}_{k}^{\prime}(\tau)$ coincide, in practice, with the poles 
of the combination $f(\rho)\, c_{eff}^2(\rho)/q(\rho)$. Recalling Eqs. (\ref{GG2})--(\ref{GG3}) we can therefore write $f(\rho)\, c_{eff}^2(\rho)/q(\rho)$ as a function of $r= \rho/\rho_{1}$, $\gamma$ and $w$:
\begin{equation}
\frac{ f(\rho) \, c_{eff}^2(\rho)}{q(\rho)} = \frac{(1 - r)}{[ 1 - (\gamma +1) r]} \, \biggl\{ c_{s,\,t}^2 + \frac{ (1+ w) r[ \gamma (\gamma +1) r - 2 \gamma]}{(1 - r)[ 1 - (\gamma +1) r]}\biggr\},
\label{GG3e}
\end{equation}
where, for the sake of illustration, $c_{s,\,t}^2 = w$; with this choice Eq. (\ref{GG3e}) applies, strictly speaking, in the case of a constant barotropic index. This is not a crucial restriction 
since Eq. (\ref{GG3e}) is always divergent in the range $0< r \leq 1$ and the 
 poles $r \to r_{\ast} \to 1/(\gamma +1)$ are not affected by the values of the barotropic index.  All in all we must then conclude that the evolution of the curvature inhomogeneities is either singular or unstable across a bounce of the scale factor. We finally stress that the gradient instabilities in the small-scale limit and the large-scale divergences are not removed by a change of the coordinate system but have instead a gauge-invariant meaning.

\subsubsection{Bounces of the curvature}
Unlike the physical situation explored in the previous subsection the bounces of the curvature are realized when $f(\rho)>0$ so that the relevant conditions can be summarized as:
\begin{equation}
f(\rho) > 0, \qquad {\mathcal H}^2 > 0, \qquad 0< \rho \leq \rho_{1}.
\label{GG5}
\end{equation}
While in the case of the bounces of the scale factor the condition (\ref{GG2}) is almost immediate (at least 
locally) various possible parametrizations may comply with Eq. (\ref{GG5}); to remain 
sufficiently simple without loosing the generality of the argument we may therefore choose:
\begin{equation}
f(r) = f_{0} \frac{r^{\gamma}}{(1 + r^2)^{\delta}},\qquad\qquad q(r) = f_{0} \frac{r^{\gamma} [ 1 + \gamma + (1 + \gamma - 2 \delta) r^2]}{(1 + r^2)^{\delta +1}},
\label{GG6}
\end{equation}
where $f_{0}$ is now a numerical constant while $\gamma >0$ and $\delta >0$ are the two shape parameters defined in the range $0< r \leq 1$. In the case of Eqs. (\ref{GG5})--(\ref{GG6}) the results of the previous 
analysis are a bit different. In particular from Eq. (\ref{GG6}) $c_{eff}^2$ is now given by:
\begin{equation}
c_{eff}^2 = c_{s,\,t}^2 + (1+w) \frac{4 \delta ^2 r^4-2 \delta  r^2 [2 \gamma 
   \left(r^2+1\right)+r^2+3]+\gamma  (\gamma +1)
   \left(r^2+1\right)^2}{\left(r^2+1\right) [\gamma +r^2 (\gamma -2 \delta
   +1)+1]}.
 \label{GG7}
 \end{equation}
 By looking at Eq. (\ref{GG7}) we can conclude that the positivity of $f(\rho)$ does not guarantee the positivity 
 of $c_{eff}^2$ and, as in the case of Fig. \ref{FIGURE1},
 the regions where $c_{eff}^2 <0$ seem generic.
\begin{figure}[!ht]
\centering
\includegraphics[height=6.7cm]{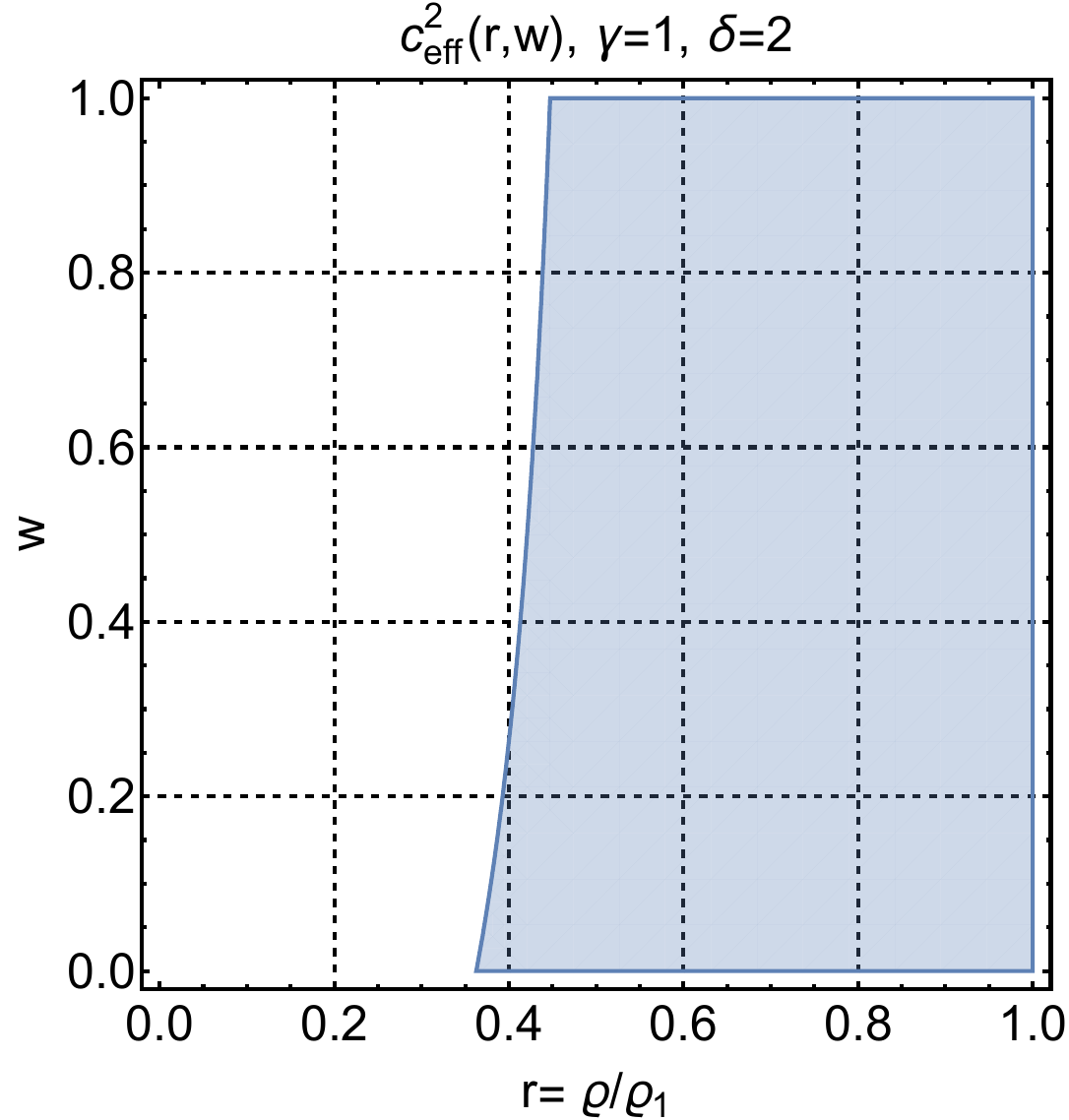}
\includegraphics[height=6.7cm]{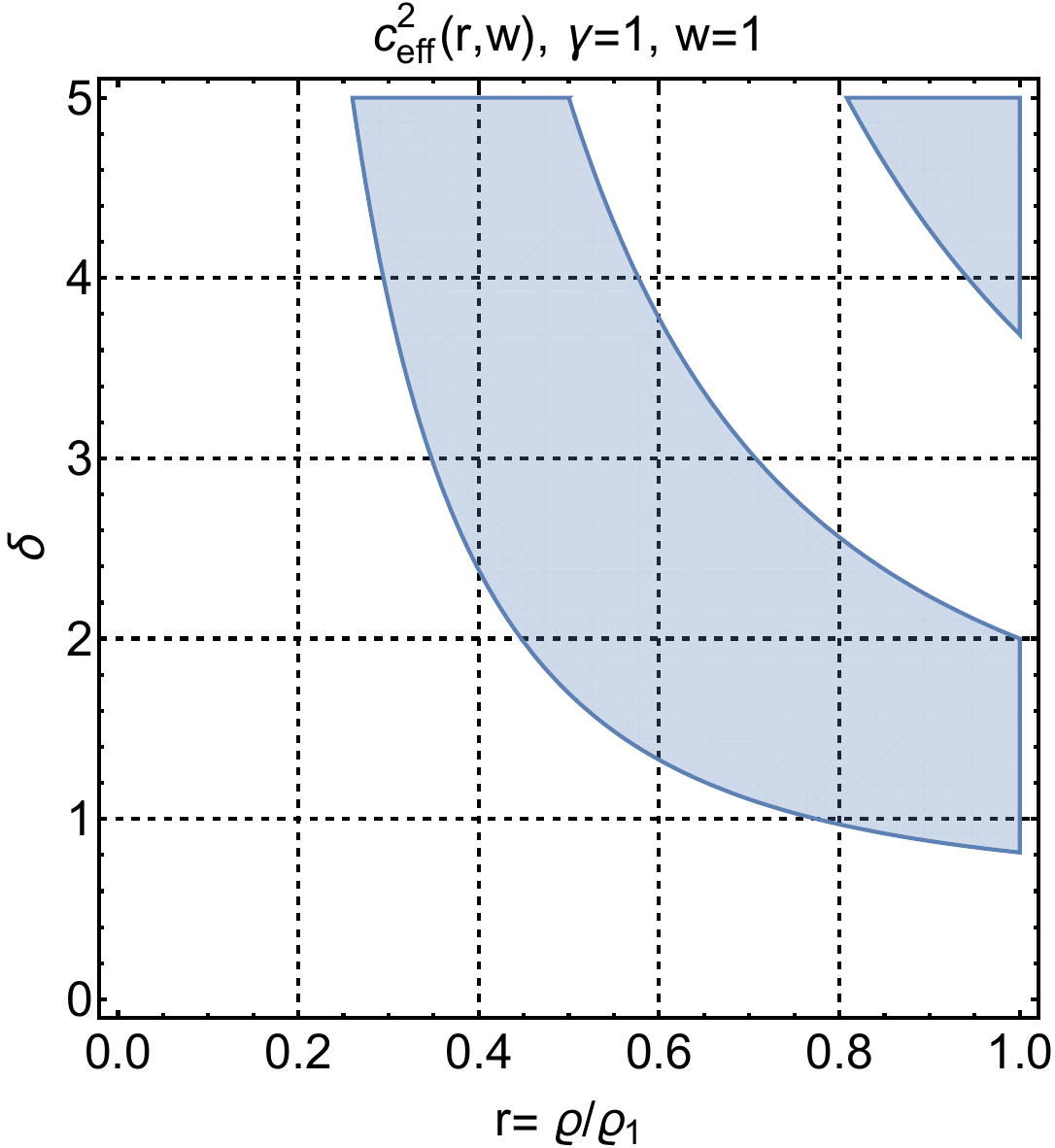}
\caption[a]{In both plots we illustrate the regions 
where $c_{eff}^2<0$ for different values of $w$ (the barotropic index), $\gamma$ and $\delta$ (see, in this respect, the parametrization of Eq. (\ref{GG6})). Both plots (valid in the case of curvature bounces) should be compared with the ones of Fig. \ref{FIGURE1} where the underlying dynamics corresponds instead to a bounce of the scale factor.}
\label{FIGURE2}      
\end{figure}
This analysis is illustrated in Fig. \ref{FIGURE2} where the shaded areas corresponds to the regions 
$c_{eff}^2 < 0$; as in the case of Fig. \ref{FIGURE1} we assumed $c_{s\,t}^2 = w$. Thus the gradient instability is not avoided in the case of curvature bounces and
the large-scale evolution of ${\mathcal R}$ also leads to singularities (see  Eq. (\ref{GG3d}) and discussion therein). We may again isolate the divergent contribution of ${\mathcal R}^{\prime}$:
\begin{equation}
\frac{f(\rho) \, c_{eff}^2(\rho)}{q(\rho)}=  \frac{m(r,\gamma, \delta, w)}{[\gamma +r^2 (\gamma -2 \delta +1)+1]^2}.
\label{GG8}
\end{equation}
In Eq. (\ref{GG8}) $m(r,\gamma, \delta, w)$ has a lengthy expression that can be written 
in terms of $3$ appropriate coefficients:
\begin{equation}
m(r,\gamma, \delta, w) = r^4 c_{1}(\gamma,\delta,w) + r^2 c_{2}(\gamma,\delta,w) + c_{3}(\gamma,\delta,w),
\label{GG9}
\end{equation}
where $c_{1}(\gamma, \delta, w)$, $c_{2}(\gamma, \delta, w)$ and $c_{3}(\gamma, \delta, w)$ are:
\begin{eqnarray}
c_{1}(\gamma,\,\delta,\,w) &=& (\gamma -2 \delta +1) [\gamma +\gamma\,w - 2 \delta  (w+1)+w],
\nonumber\\
c_{2}(\gamma,\,\delta,\,w) &=& 2 (\gamma +1) [\gamma + w (\gamma +1)] -2 \delta  [2 \gamma +2 (\gamma +2) w+3],
\nonumber\\
c_{3}(\gamma,\,\delta,\,w) &=& (\gamma +1) [\gamma + w (\gamma +1)].
\label{GG10}
\end{eqnarray}
 By looking at Eqs. (\ref{GG8})--(\ref{GG9}) and (\ref{GG10}) we see that the only potential divergences may occur when 
 $\gamma < 2 \delta -1$: only in this case the denominator appearing at the right-hand side of Eq. (\ref{GG8}) may vanish for a real value of $r$ since, by definition, $\gamma +1>0$. We can therefore evaluate $m(r, \gamma, \delta, w)$ at the poles; from the denominator of Eq. (\ref{GG8}) 
 we see that the  potential singularities correspond to $r_{\ast}^2$:
 \begin{equation}
 r_{\ast}^2 = \frac{1 + \gamma}{2 \delta - (\gamma -1)}, \qquad \qquad m(r_{\ast}, \gamma, \delta, w) = \frac{4 (\gamma +1) \delta  (w+1)}{\gamma -2 \delta +1}.
 \label{GG11}
 \end{equation}
When $\gamma < 2 \delta -1$ the points $r_{\ast}$ correspond then to the potential divergences 
 of the curvature inhomogeneities. This observation suggests that the curvature inhomogeneities are 
 probably finite when $\gamma > 2 \delta -1$. For this reason we then conclude that the degree of divergence of ${\mathcal R}^{\prime}$ is less severe in comparison with the bounces of the scale factor treated in the previous subsection.
 
\subsection{Scalar field matter}
So far we investigated the case of perfect and irrotational fluids and  
in the presence of scalar field matter the conclusions are conceptually similar but quantitatively different.
The reason for this difference stems directly from the fluctuations of the energy momentum tensor of Eq. (\ref{DD4}) since, in the scalar field case, $\delta_{s} T_{\mu}^{\,\,\,\nu}$ takes a different form that can be covariantly written as:
\begin{eqnarray}
\delta_{s} \, T_{\mu}^{\,\,\nu} &=& \frac{\partial}{\partial\rho_{\varphi}} \biggl[ \rho_{\varphi} \, f(\rho_{\varphi})\biggr] \delta_{s} \rho_{\varphi} \,\, \delta_{\mu}^{\,\,\nu} + \frac{\partial q}{\partial \rho_{\varphi}} \biggl[ - (p_{\varphi} + \rho_{\varphi}) \delta_{\mu}^{\,\, \nu} + \partial_{\mu} \varphi \,\partial^{\nu} \varphi\biggr] \delta_{s} \rho_{\varphi}
\nonumber\\
&+& q (\rho_{\varphi}) \biggl[ \partial_{\mu} \varphi \,\partial^{\nu} \chi + \partial_{\mu} \chi \,\partial^{\nu} \varphi + \delta_{s} g^{\nu\alpha}\partial_{\alpha} \varphi\,\partial_{\mu}\varphi - \delta_{\mu}^{\nu} \biggl(\delta_{s} \rho_{\varphi} + \delta_{s} p_{\varphi}\biggr)\biggr],
\label{HH1}
\end{eqnarray} 
where, by definition, $\delta_{s} \varphi = \chi$ is the scalar field fluctuation. Equation (\ref{HH1}) holds in general terms and, in the gauge (\ref{FF1}),
the different components of $\delta_{s} \, T_{\mu}^{\, \nu}$ are given explicitly by: 
\begin{eqnarray}
\delta_{s} T_{0}^{\,\,0} &=& \frac{\partial}{\partial\rho_{\varphi}}\bigl[ \rho_{\varphi} \, f(\rho_{\varphi})\bigr] \, \delta_{s} \rho_{\varphi}
\label{HH2}\\
\delta_{s} T_{i}^{\,\,j} &=& \biggl\{ \frac{\partial}{\partial\rho_{\varphi}} \biggl[ \rho_{\varphi} \, f(\rho_{\varphi})\biggr] \delta_{s} \rho_{\varphi}
- \frac{\partial q}{\partial \rho_{\varphi}} ( p_{\varphi} + \rho_{\varphi}) \delta_{s} \rho_{\varphi} - q(\rho_{\varphi}) \bigl[ \delta_{s} p_{\varphi} + 
\delta_{s} \rho_{\varphi}\bigr]\bigr\} \delta_{i}^{\,\,j},
\label{HH3}\\
\delta_{s} \, T_{i}^{\,\,0} &=& q(\rho_{\varphi}) \biggl(\frac{\varphi^{\prime}}{a^2}\biggr) \partial_{i} \chi, 
\label{HH4}\\
\delta_{s} \, T_{0}^{\,\,i} &=& - \frac{q}{a^2} \biggl[ \varphi^{\prime} \partial^{i} \chi + \varphi^{\prime\, 2} \, \partial^{i} B \biggr].
\label{HH5}
\end{eqnarray}
In Eqs. (\ref{HH2})--(\ref{HH3}) two auxiliary quantities have been introduced, namely $\delta_{s}\rho_{\varphi}$ and $\delta_{s} p_{\varphi}$, i.e. the scalar fluctuations of the energy density and of the scalar pressure:
\begin{eqnarray}
\delta_{s} \rho_{\varphi} &=& \frac{1}{a^2} \biggl( - \phi\, \varphi^{\prime\, 2} + \varphi^{\prime} \chi^{\prime} + \frac{\partial V}{\partial\varphi} a^2 \chi\biggr),
\label{HH6}\\
\delta_{s} p_{\varphi} &=& \frac{1}{a^2} \biggl( - \phi\, \varphi^{\prime\, 2} + \varphi^{\prime} \chi^{\prime} - \frac{\partial V}{\partial\varphi} a^2 \chi\biggr).
\label{HH7}
\end{eqnarray}
To derive the evolution of ${\mathcal R}$ in this case we may first analyze the momentum constraint coming from the $(0\, i)$ components of the perturbed Einstein equations, namely 
\begin{equation}
 \delta_{s} {\mathcal G}_{i}^{\,\,0} = \ell_{P}^2 \delta_{s} T_{i}^{\,\,0} \qquad \Rightarrow \qquad  2 {\mathcal H} \phi =   \ell_{P}^2 \, q \, \varphi^{\prime} \chi,
\label{HH8}
\end{equation}
where we employed Eqs. (\ref{HH4}) and (\ref{APPA3}). If combined with the expression of the curvature inhomogeneities defined in Eq. (\ref{FF3}) we can obtain the direct
connection between ${\mathcal R}$ and $\chi$:
\begin{equation}
{\mathcal R} = - \frac{{\mathcal H}^2 }{{\mathcal H}^2 - {\mathcal H}^{\prime}} \phi = - \biggl(\frac{{\mathcal H}}{\varphi^{\prime}}\biggr) \,\chi.
\label{HH9}
\end{equation}
Note that $({\mathcal H}^2 - {\mathcal H}^{\prime})$ and ${\mathcal H}\, \phi$ contain the same dependence on $q(\rho_{\varphi})$: this is why in the second equality of Eq. (\ref{HH9}) $q(\rho_{\varphi})$ disappears. The Hamiltonian constraint stemming from the $(0\,0)$ component of the perturbed Einstein's equation is finally given by:
\begin{eqnarray}
{\mathcal H} \nabla^2 B + 3 {\mathcal H}^2 \phi + \frac{\ell_{P}^2}{2} {\mathcal q}(\rho_{\varphi})\biggl( \chi^{\prime} \varphi^{\prime} + \frac{\partial V}{\partial \varphi} a^2 \chi \biggr) =0.
\label{HH10}
\end{eqnarray}
For immediate convenience, in Eq. (\ref{HH10}) we introduced the combination:
\begin{equation}
{\mathcal q}(\rho_{\varphi}) = \biggl[ f + \rho_{\varphi} \frac{\partial f}{\partial \rho_{\varphi}}\biggr].
\label{HH10a}
\end{equation}
Although ${\mathcal q}(\rho_{\varphi})$ and $q(\rho_{\varphi})$ ultimately coincide if the standard 
form of the covariant conservation is enforced (see Eq. (\ref{DD6})), it is useful to keep them formally distinct so that, in this way, the obtained results may also cover the limit $q(\rho_{\varphi}) \neq {\mathcal q}(\rho_{\varphi})$.
Since, thanks to Eq. (\ref{FF3}), $\phi$ is proportional to ${\mathcal R}$ and  $\chi$ is proportional to ${\mathcal R}$ (because of Eq. (\ref{HH9})), $\phi$ and $\chi$ can be traded for ${\mathcal R}$ in Eq. (\ref{HH10}). The  result of this manipulation leads to  the scalar field analog of Eq. (\ref{FF14})
\begin{equation}
{\mathcal R}^{\prime} = \frac{{\mathcal H}^2 }{4 \pi G \overline{q} \varphi^{\prime\, 2}} - 3 {\mathcal H} \biggl[ 1 - \frac{q(\rho_{\varphi})}{{\mathcal q}(\rho_{\varphi})}\biggr] {\mathcal R},
\label{HH11}
\end{equation}
and if we now enforce the standard form of the covariant conservation (as implied by Eq. (\ref{DD6})) we obtain, as anticipated, that Eq. (\ref{HH11}) becomes: 
\begin{equation}
{\mathcal R}^{\prime} =  \frac{{\mathcal H}^2}{ 4 \pi G \, {\mathcal q}\,  \varphi^{\prime\, 2}} \nabla^2 B = - \frac{{\mathcal H}}{ 4 \pi G q \varphi^{\prime\, 2}} \nabla^2 \Psi.
\label{HH12}
\end{equation}
Recalling the discussion of Eq. (\ref{APPA4}), the second equality follows from the observation that, in the gauge (\ref{FF1}) $\Psi = - {\mathcal H} B$. 

Concerning the comparison between Eqs. (\ref{FF14}) and (\ref{HH11}) two comments are in order. While in Eq. (\ref{FF14}) there appears the effective sound speed, a similar term does not show up in Eq. (\ref{HH12}).
The second remark is that the term ${\mathcal S}_{{\mathcal R}}$ (containing the anisotropic stress and the non-adiabatic pressure fluctuation) has no analog at least in the case when ${\mathcal q} \to q$.
Bearing in mind these two point,  the derivation follows the same steps already outlined in the fluid case. 
In particular we first rephrase Eq. (\ref{HH12}) as 
\begin{equation}
{\mathcal R}^{\prime} = \frac{1}{4 \pi G} \biggl(\frac{a^2}{q z^2}\biggr)  \nabla^2 B, \qquad \qquad z = \frac{a\, \varphi^{\prime}}{{\mathcal H}}.
\label{HH13}
\end{equation}
We must then compute the time derivative of both sides of Eq. (\ref{HH13}) and then eliminate $B^{\prime}$ by using the relation coming from the spatial component of the perturbed Einstein's equations, i.e. 
\begin{equation}
\phi + B^{\prime} + 2 {\mathcal H} B=0. 
\end{equation}
At the end of this procedure the following equation for ${\mathcal R}$ is obtained:
\begin{equation}
{\mathcal R}^{\prime\prime} + \biggl(2 \frac{z^{\prime}}{z} + \frac{q^{\prime}}{q}\biggr) {\mathcal R}^{\prime} - \nabla^2 {\mathcal R} =0.
\label{HH14}
\end{equation}
Equation (\ref{HH14}) can be phrased in an even simpler form:
\begin{equation} 
 {\mathcal R}^{\prime\prime} + 2 \frac{z_{\varphi}^{\prime}}{z_{\varphi}}  {\mathcal R}^{\prime} - \nabla^2 {\mathcal R} =0, \qquad \qquad z_{\varphi}^2 = z^2\, q\, = \frac{a^2 \, \varphi^{\prime\, 2} q}{{\mathcal H}^2}.
\label{HH15}
\end{equation}
In the limit $q \to 1$ Eq. (\ref{HH15}) reproduces the standard evolution equation of curvature inhomogeneities 
in the case of a single scalar field. We are now going to analyze, in general terms, the singularity properties of 
Eq. (\ref{HH15}) in  analogy with what already discussed in the case of the irrotational fluid.

\subsubsection{Bounces of the scale factor}
Although ${\mathcal R}^{\prime}$is not generally regular, the main difference in comparison with 
the fluid case is related to the absence of the effective sound speed (i.e. $c_{eff}^2 \to 1$ in the scalar case).  
 As in the fluid case however we may transform the differential equation 
(\ref{HH15}) into an integral equation and integrating once Eq. (\ref{HH15}) we obtain: 
\begin{equation}
{\mathcal R}_{k}^{\prime}(\tau) = {\mathcal R}_{k}^{\prime}(\tau_{ex})\, \frac{z_{ex}^2}{z_{\varphi}^2(\tau)} - 
\frac{k^2}{ z_{\varphi}^2(\tau)} \int_{\tau_{ex}}^{\tau} z_{\varphi}^2(\tau_{1}) \,\, {\mathcal R}_{k}(\tau_{1}) \, d\tau_{1}.
\label{HH15a}
\end{equation}
Equation (\ref{HH15a}) must be further integrated so that we get an expression which is analog to the one already derived in Eqs. (\ref{GG3c})--(\ref{GG3d}):
\begin{eqnarray}
{\mathcal R}_{k}(\tau) &=& {\mathcal R}_{k}(\tau_{ex}) + {\mathcal R}_{k}^{\prime}(\tau_{ex}) \, \int_{\tau_{ex}}^{\tau} \frac{z_{ex}^2}{z_{\varphi}^2(\tau_{1})}\,\, d\tau_{1}
\nonumber\\
&-& k^2 \, \int_{\tau_{ex}}^{\tau} \frac{d \tau_{1}}{z_{\varphi}^2(\tau_{1})} \, \int_{\tau_{ex}}^{\tau} \, z_{\varphi}^2(\tau_{2}) \, {\mathcal R}_{k}(\tau_{2})\, d\tau_{2}.
\label{HH15b}
\end{eqnarray}
Finally Eq. (\ref{HH15b}) should be iteratively solved and, to leading order in $k^2 \tau^2 \ll 1$,
the divergences of ${\mathcal R}^{\prime}(\tau)$ correspond indeed  to the zeros of $z_{\varphi}^2$:
\begin{equation}
z_{\varphi}^2 = \frac{a^2 \varphi^{\prime \, 2}}{{\mathcal H}^2} q(\rho_{\varphi})= \frac{3 \overline{M}_{P}^2\, a^4}{(1 + 2 \, a^2\, V/\varphi^{\prime\, 2})}  \,\, 
\frac{q(\rho_{\varphi})}{f(\rho_{\varphi})}.
\label{HH15c}
\end{equation}
In the case of a regular solution describing a bounce of the scale factor the  zeros 
of $z_{\varphi}^2(\tau)$ must inevitably coincide with the zeros of $q(\rho_{\varphi})/f(\rho_{\varphi})$. Thus the singularities of ${\mathcal R}_{k}^{\prime}(\tau)$ are in fact 
the poles of $f(\rho_{\varphi})/q(\rho_{\varphi})$, namely
\begin{equation}
\frac{f(r, \gamma)}{q(r,\gamma)} = \frac{r -1}{r\, (\gamma +1) -1}.
\label{HH15d}
\end{equation}
Equation (\ref{HH15d}) implies, as expected, that the singularities in ${\mathcal R}_{k}^{\prime}(\tau)$ are generic as long as $f(\rho)$ contains a zero of order $\gamma$ for $0< r \leq 1$. In other words, if we assume that $f(\rho_{\varphi})$ has a zero then the dynamical evolution 
of the system always implies (at least) a second zero of $z_{\varphi}^2$ (and hence a divergence 
of the curvature inhomogeneities).

\subsubsection{Bounces of the curvature}
For the bounces of the curvature the discussion is qualitatively similar but also physically different. 
As already mentioned, in this case $f(\rho_{\varphi}) >0$; we can tentatively parametrize $f(\rho_{\varphi})$ 
in terms of a positive-definite function without poles or zeros. For example 
a sound choice could be the one already suggested in Eq. (\ref{GG6}) with the difference 
that now $\rho$ must be replaced by $\rho_{\varphi}$
\begin{equation}
f(\rho_{\varphi}) = f_{0} \frac{r^{\gamma}}{( 1 + r^2 )^\delta}, \qquad \qquad r= \rho_{\varphi}/\rho_{1}.
\label{HH20}
\end{equation}
In this case we can repeat the previous considerations and discover that, again, the zeros of $q(\rho_{\varphi})/f(\rho_{\varphi})$
correspond to the potential singularities of ${\mathcal R}^{\prime}$. If we then use  Eq. (\ref{HH20}) 
we obtain 
\begin{equation}
\frac{f(r, \gamma)}{q(r,\gamma)}  = \frac{r^2+1}{(\gamma +1) +r^2 (\gamma -2 \delta +1)}.
\label{HH21}
\end{equation}
Since the expression of Eq. (\ref{HH21}) does not go to zero, there are chances that this class 
of curvature bounces lead to a regular evolution of curvature inhomogeneities when $(\gamma - 2 \delta + 1)>0$.

\subsection{Background-independent expectations}
If the effective energy and enthalpy densities are modified with two smearing functions 
(eventually constrained by the covariant conservation of the total energy momentum tensor)
the evolution of the curvature inhomogeneities presents some general features that can be 
analyzed in a background-independent manner. The results obtained so far can be summarized, in short, as follows.
\begin{itemize}
\item{} When the energy momentum tensor is modelled in terms of a perfect irrotational fluid, the 
presence of the smearing functions induces an effective sound speed in the evolution of curvature inhomogeneities. The effective sound speed is not positive definite but even assuming this crucial requirement the evolution of ${\mathcal R}^{\prime}$ is not necessarily regular in the large-scale limit.
\item{} When the energy-momentum tensor is dominated by a scalar field the potential  instability appearing in the fluid case does not occur. Still the evolution of ${\mathcal R}^{\prime}$ may diverge for the bounces 
of the scale factor. For curvature bounces it may be possible to avoid instabilities and divergences, at least in some very specific classes of scenarios.
\end{itemize}
\renewcommand{\theequation}{4.\arabic{equation}}
\setcounter{equation}{0}
\section{Explicit examples and applications}
\label{sec4}
Although the previous considerations made use of the conformal time coordinate, the concrete analysis of various explicit solutions becomes more transparent in the cosmic time and a swift dictionary between the two descriptions has been collected in appendix \ref{APPB} with particular attention to the evolution of the curvature inhomogeneities. The coordinate systems are in fact matter of convenience 
and the results of the previous sections do not depend on the time parametrization.
With these caveats, in cosmic time Eqs. (\ref{CC6})--(\ref{CC7}) reduce to 
\begin{eqnarray}
3 H^2 \,\overline{M}_{P}^2 = \rho \, f(\rho),\qquad\qquad
2 \dot{H} \, \overline{M}_{P}^2 = - (p + \rho) q(\rho), 
 \label{LL1}
 \end{eqnarray}
 where we introduced the Planck mass $\overline{M}_{P} = 1/\ell_{P}$. In Eq. (\ref{LL1}) the 
 overdot denotes, as usual, a derivation with respect to the cosmic time coordinate and $ H = \dot{a}/a$. 
 In this section $q(\rho)$ and $f(\rho)$ are related as in Eq. (\ref{CC2}); in other words, both for the perfect fluid and and for the scalar field matter we have:
 \begin{equation}
q(\rho) = f(\rho) + \rho\, \frac{\partial f}{\partial \rho}, \qquad q(\rho_{\varphi}) = f(\rho_{\varphi}) + \rho_{\varphi}\, \frac{\partial f}{\partial \rho_{\varphi}}.
\label{LL2}
\end{equation}
For the scalar field matter Eq. (\ref{LL1}) holds, in practice, by replacing $\rho$ and $p$ with 
$\rho_{\varphi}$ and $p_{\varphi}$; the relevant equations become then: 
\begin{eqnarray}
3 H^2 \,\overline{M}_{P}^2 = \rho_{\varphi} \, f(\rho_{\varphi}),\qquad\qquad
2 \dot{H} \, \overline{M}_{P}^2 = - (p_{\varphi} + \rho_{\varphi}) q(\rho_{\varphi}),
 \label{LL3}
 \end{eqnarray}
where $\rho_{\varphi}$ and $p_{\varphi}$ are given by Eq. (\ref{CC9}) but expressed in the 
cosmic time parametrization 
\begin{equation}
\rho_{\varphi} = \frac{\dot{\varphi}^2}{2} + V(\varphi), \qquad p_{\varphi} = \frac{\dot{\varphi}^2}{2} - V(\varphi).
\label{LL4}
\end{equation}
Equation (\ref{LL2}) implies that for the perfect fluid case and for the scalar field matter 
the standard forms of the continuity and of the Klein-Gordon equations hold:
\begin{equation}
\dot{\rho} + 3 H (\rho + p) =0, \qquad\qquad \ddot{\varphi} + 3 H \dot{\varphi} + \frac{\partial V}{\partial \varphi} =0.
\label{LL5}
\end{equation}
Some specific solutions will now be analyzed with the purpose of corroborating 
the perspective presented in the previous sections.

\subsection{Bounces of the scale factor}
\subsubsection{Perfect irrotational fluids} 
\begin{figure}[!ht]
\centering
\includegraphics[height=6.2cm]{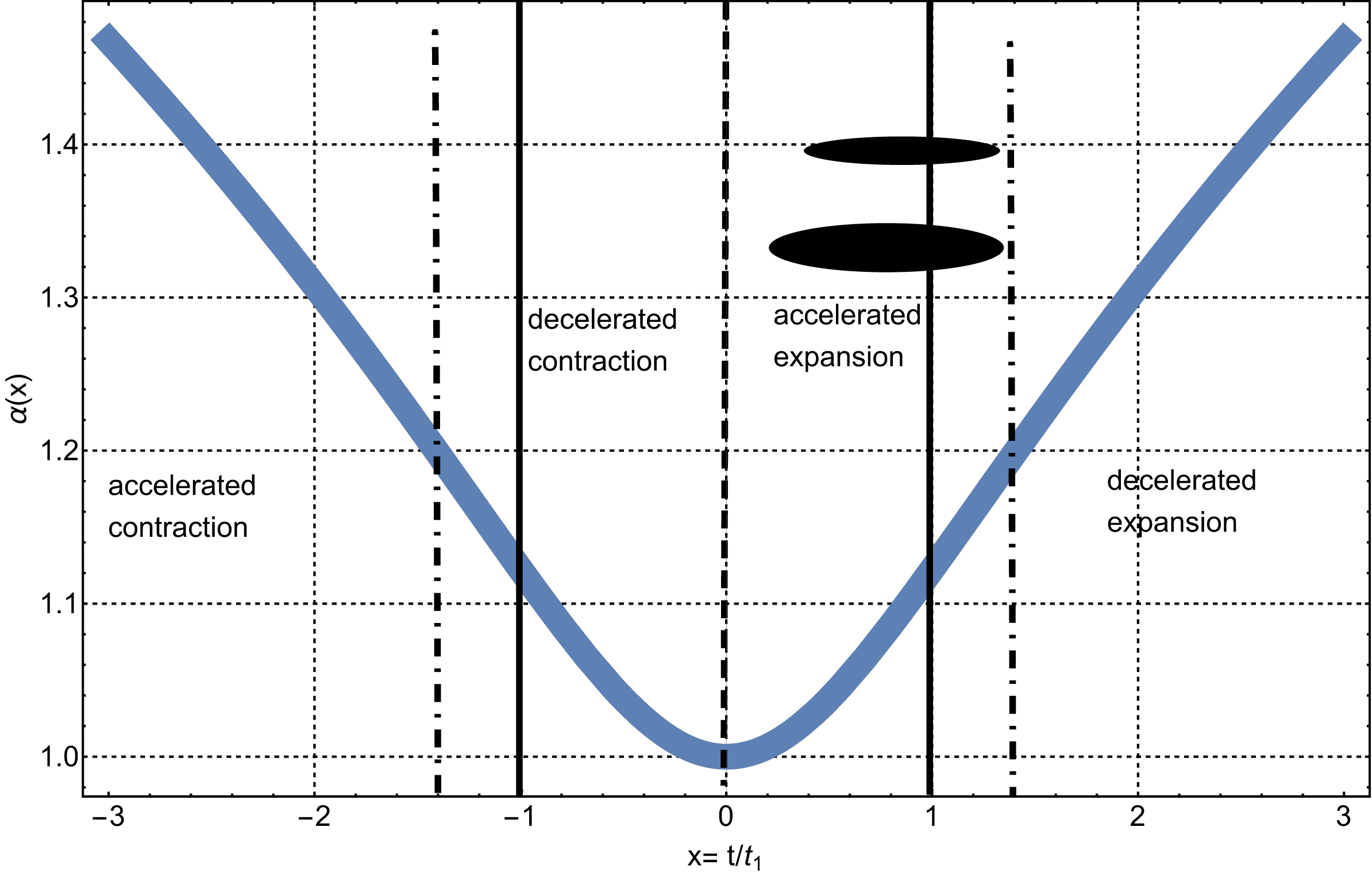}
\caption[a]{We illustrate here the different kinematical regions of the bouncing solutions associated with Eq. (\ref{LL8}). In particular this plot has been obtained in the case $w=1$ but the same considerations hold indifferently also for other  bounces of the scale factor discussed throughout this section. The vertical (dashed) line separates the contracting stage (occurring for $t <0$) from the expanding phase (for $t>0$). The dot-dashed 
line at the left divides the epoch of accelerated contraction from the one of decelerated contraction (always 
taking place for $t <0$). The vertical dot-dashed line at the right separates instead the phase of accelerated expansion from the one of decelerated expansion. Finally, the two full (vertical) lines illustrate the divergences arising in the evolution of the curvature inhomogeneities for the large-scale limit. The ellipses correspond approximately to the regions where one would be tempted to set the initial conditions of the subsequent evolution. These regions are however excluded by the  divergences of the curvature inhomogeneities.}
\label{FIGUREDEF1}      
\end{figure}

According to Eq. (\ref{GG2}) the  bounces of the scale factor are characterized by a zero of $f(\rho)$
which may also lead to a zero in $q(\rho)$ (even if not for the same value of the energy density). Both zeros do not affect the regular character of the solution even though an instability is expected 
according to the general arguments of section \ref{sec3}.
Let us therefore consider the simplest possibility, namely the case $\gamma=1$ in Eq. (\ref{GG2}) and see if the concrete solution obtainable in this case really leads to the background-independent 
conclusions deduced in the previous section. From the continuity equation (and in the case of a perfect irrotational fluid) we have that 
\begin{equation}
\rho = \rho_{1} (a_{1}/a)^{3(w +1)} \qquad \Rightarrow\qquad r(\alpha) = \alpha^{- 3 (w+1)}, \qquad  \alpha = (a/a_{1}).
\label{LL6}
\end{equation}
The Hubble rate follows from Eq. (\ref{LL1}) written in the specific case of Eq. (\ref{LL6})
\begin{equation}
3\, H^2 \, \overline{M}_{P}^2 = \rho_{1} \alpha^{-3 (w+ 1)} \bigl( 1 - \alpha^{-3(w+1)}),
\label{LL7}
\end{equation}
implying that the solution of the whole system for $\gamma \to 1$ can be written in the form 
\begin{equation}
\alpha(x) = (x^2 +1)^{\sigma}\,\, \Rightarrow\,\, H(x) = H_{1}\,\,\frac{x}{x^2+1}, \qquad x = t/t_{1}, \qquad 
\sigma = \frac{1}{3(w + 1)},
\label{LL8}
\end{equation}
where $H_{1} = 2\sigma/t_{1}$ and $\rho_{1} = 3 \, H_{1}^2 \, \overline{M}_{P}^2$.
The  evolution suggested by Eq. (\ref{LL8}) follows by considering separately the first and second time derivatives of the scale factor:
\begin{eqnarray}
\dot{a}(x) &=& a_{1}\, H_{1}\, \,x \,\frac{\alpha(x)}{x^2 + 1},
\label{LL8b}\\
\ddot{a}(x) &=& \frac{3 (w +1)}{2} \, a_{1} \, H_{1}^2 \frac{\alpha(x)}{(x^2 +1)^2} \biggl[ 1 - \frac{(1 + 3 w)}{3 ( w+1)} x^2\biggr].
\label{LL8c}
\end{eqnarray}
From Eq. (\ref{LL8b}) we have that $\dot{a} >0$ (for $t>0$) and $\dot{a} <0$ (for $t<0$).
Equation (\ref{LL8c}) implies instead a more complicated series of conditions:
\begin{eqnarray}
&&\ddot{a} > 0, \qquad \mathrm{for}  \qquad  - t_{\ast}  < \, t \, < t_{\ast}, 
\nonumber\\
&& \ddot{a}<0, \qquad \mathrm{for} \qquad t <  - t_{\ast} \, \qquad \mathrm{and\,\,\,\,for}\qquad 
 t > t_{\ast},
 \label{LL8d}
\end{eqnarray}
where $t_{\ast} = \pm \, t_{1} \, \sqrt{3 (w+1)/(3 w +1)}$. Equations (\ref{LL8b})--(\ref{LL8c}) and 
(\ref{LL8d}) lead, therefore, to the timeline illustrated in the cartoon of Fig. \ref{FIGUREDEF1}
where the times $\pm t_{\ast}$ are marked  by a dot-dashed line.
For $t < - t_{\ast}$ an accelerated contraction (i.e. $\dot{a} <0$ and $\ddot{a} <0$) is followed 
by a decelerated expansion ($\dot{a} >0$ and $\ddot{a} <0$) in the region $t > t_{\ast}$. There are, however, two further intermediate regions: for $- \, t_{\ast} < t <0$  a decelerated contraction (i.e. $\dot{a} <0$ and $\ddot{a} >0$) is followed by a (short) inflationary stage (i.e. $\dot{a} >0$ and $\ddot{a} >0$) in the range $0 < t < t_{\ast}$.

The solution of Eq. (\ref{LL8}) (or some analog solution) could then be used 
to set initial conditions for a putative inflationary stage occurring much later. 
Indeed Eq. (\ref{LL8}) only assumes the presence of the scalar kinetic energy and the 
inclusion of a scalar potential at late time could lead to the wanted inflationary stage.
According to this picture the initial conditions could then be set when the kinetic energy is still dominant. We see however from Fig. \ref{FIGUREDEF1} that the only viable range where the initial conditions could be eventually 
imposed is the one for $t \gg t_{1}$ (i.e. $x \gg 1$). The full vertical lines $x = \pm 1$ 
in  Fig. \ref{FIGUREDEF1} correspond in fact to singularities in the evolution 
of the curvature inhomogeneities: if  the initial conditions 
are set for $t < t_{1}$ the subsequent evolution will go through a potentially unstable region 
where the large-scale curvature inhomogeneities diverge.
We remark that the solutions described here look similar to the ones employed, in a related 
context, by the authors of Refs. \cite{DD1,DD2}.
The elliptic regions appearing in Fig. \ref{FIGUREDEF1} illustrate,  approximately, the range where the authors of Refs. \cite{DD1,DD2} decided to set their initial conditions in the context of a solution obtained, from the present viewpoint,  from a specific choice of $f(\rho)$ and $q(\rho)$.
While it is claimed that this solution could eventually solve the problem of the initial singularity,
the general arguments present here suggest, however, that the 
divergences in $\dot{{\mathcal R}}$ may severely restrict the dynamical range of the solution, as we are going to analyze even more explicitly in the following subsection.
 
 \subsubsection{Curvature inhomogeneities}
\begin{figure}[!ht]
\centering
\includegraphics[height=5.5cm]{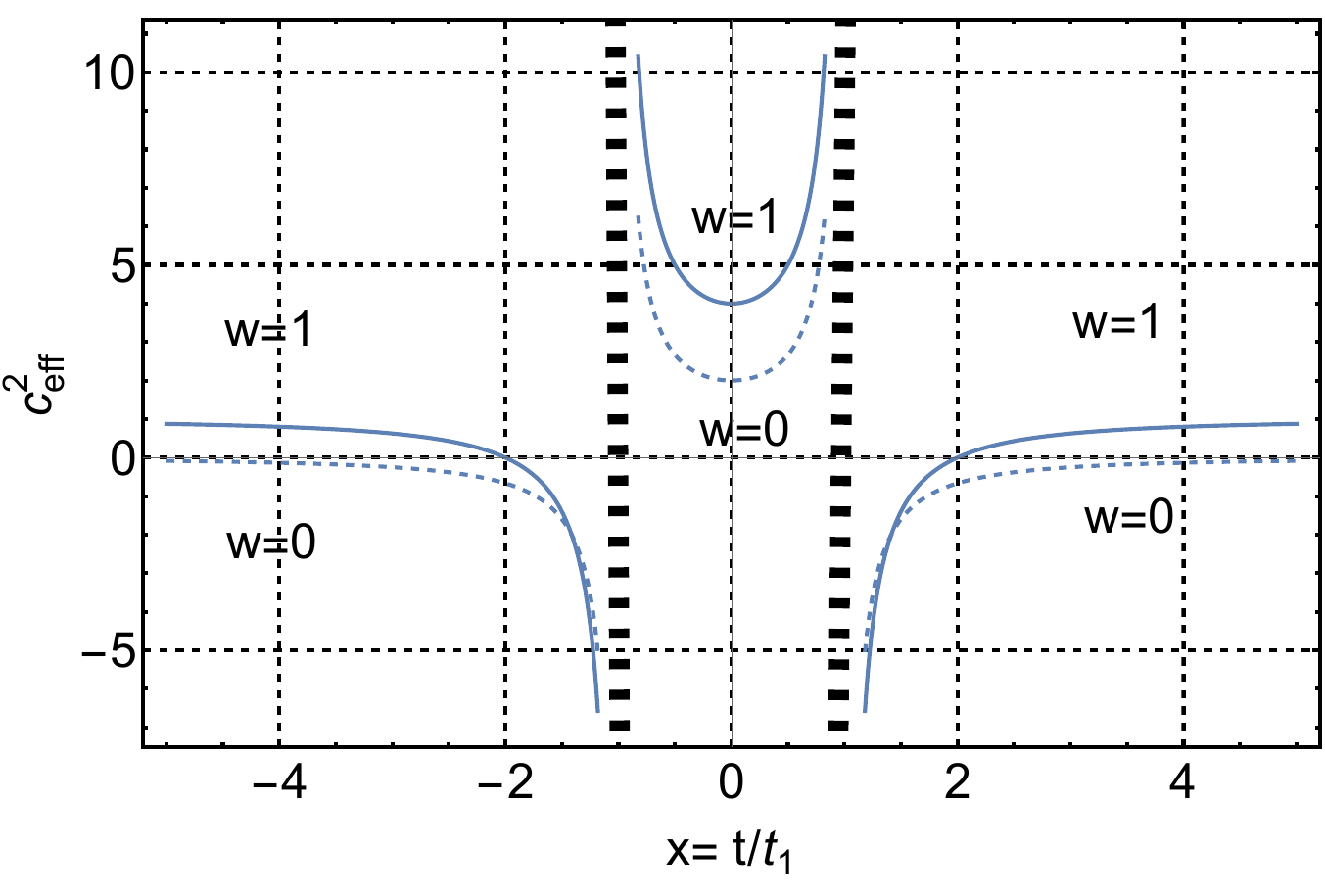}
\includegraphics[height=5.5cm]{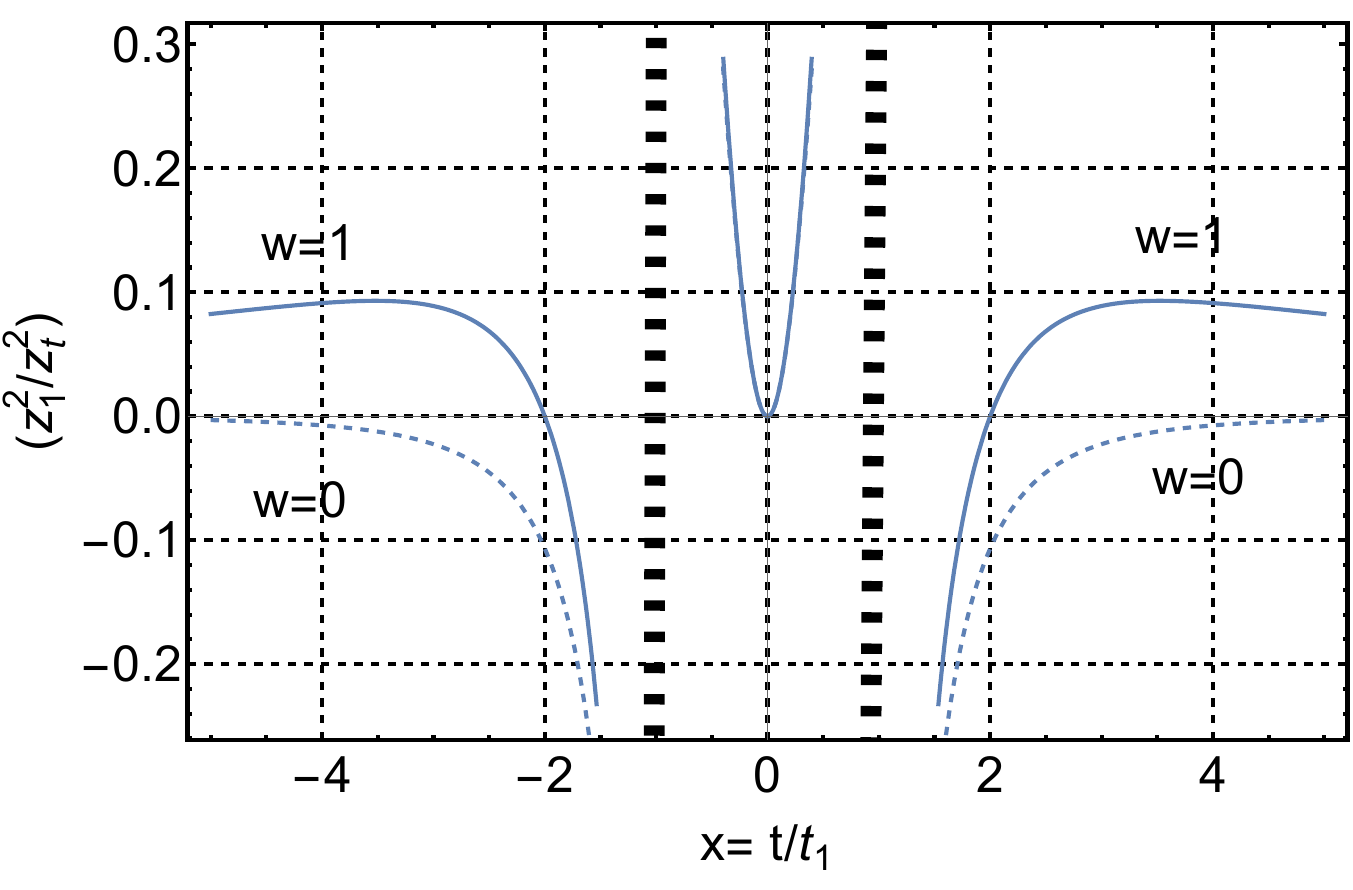}
\caption[a]{In this example (see plot at the left) $c_{eff}^2(x)$ is positive definite for $| x|<1$ while $c_{eff}^2(x)$ gets negative in the region $|x| > 1$. Between the two ranges the effective sound speed diverges when $ x  \to \pm 1$. In the right plot we illustrate instead $(z_{1}/z_{t})^2$ always  as a function of the rescaled time coordinate (i.e. $x=t/t_{1}$) and for two different values of $w$. The regions where $c_{eff}^2(x)<0$ signal 
in practice the small-scale instabilities while the singularities appearing in the right plot correspond to the singularities of $\dot{{\mathcal R}}$ in the 
large-scale limit.}
\label{FIGURE3}      
\end{figure}

 In Eq. (\ref{GG3}) we demonstrated, in general terms, that  the square of the 
effective sound speed may get negative in a wide portion of the evolution. We can now compute explicitly $c_{eff}^2(x)$ in the 
specific background of Eq. (\ref{LL8}):
\begin{equation}
c_{eff}^2(\alpha) = \frac{w - 2 r(\alpha)( 2\, w +1)}{ 1 - 2 r(\alpha)}, \qquad c_{s,\,t}^2 = w.
\label{LL9}
\end{equation}
In terms of the cosmic time coordinate $c_{eff}^2(x)$ does not have a definite sign and the singularities appearing as a function 
of $r$ also show up in terms of $x = t/t_{1}$:
\begin{equation}
c_{eff}^2(x) = \frac{w\, x^2 - (3 w+2)}{x^2-1}.
\label{LL10}
\end{equation}
 In the left plot of Fig. \ref{FIGURE3} the profile of $c_{eff}^2(x)$ is illustrated for 
two different values of $w$. 
The results illustrated in Fig. \ref{FIGURE3} are particularly unphysical since $c_{eff}^2 < 0$  far from the bounce even if in the asymptotic regions (where the background either contracts or expands)
the evolution of curvature inhomogeneities may be stable as long as $c_{eff}^2 \to w$. The potential instabilities involve, in particular, the spatial gradients which enter the evolution of ${\mathcal R}$ with the wrong sign. We may actually go back to Eq. (\ref{FF23}) written in the cosmic time parametrization 
\begin{equation}
\ddot{{\mathcal R}}_{k} + 2 \frac{\dot{z}_{t}}{z_{t}} \dot{{\mathcal R}}_{k} + \frac{k^2\,c_{eff}^2(t)}{a^2(t)} {\mathcal R}_{k} =0,  \qquad z_{t}^2(t) = \frac{a^{3} \, (p + \rho)\, q(\rho)}{H^2 \, c_{eff}^2}.
\label{LL11}
\end{equation}
Equation (\ref{LL1}) follows from Eq. (\ref{FF23}) by taking into account the original form of the pump field and by 
rephrasing it in the cosmic time parametrization (see appendix \ref{APPB} and discussion therein).
In the case of Eq. (\ref{LL11}) the zeros of $z_{t}^{2}$ correspond to the divergences of $\dot{\mathcal R}$
which is proportional to $z_{t}^{-2}$. From Eqs. (\ref{LL8}) and (\ref{LL9})--(\ref{LL10}) we have  
\begin{equation}
z_{t}^2(x) =  z_{1}^2 \,(1+w) \, \frac{(x^2 -1)\, (x^2 +1)^{(w+2)/(w+1)}}{x^2 \, [ w (x^2 +1) - (3 w+2)]}, \qquad\qquad
z_{1}^2 = \frac{a_{1}^3 \rho_{1}}{H_{1}^2}.
\label{LL12}
\end{equation}
The inverse of Eq. (\ref{LL12}) is illustrated in the right plot of Fig. \ref{FIGURE3} for two different 
values of $w$; in spite of the value of $w$ the divergences of $\dot{{\mathcal R}}_{k}$ are always 
present for $x=\pm1$. Furthermore in the central region of the right plot of Fig. \ref{FIGURE3} (i.e. $ |x| < 1$) the curves corresponding to different values of $w$ are indistinguishable and this happens because for $|x|< 1$ the leading-order result of the expansion does not depend on $w$, i.e. $(z_{1}/z_{t})^2 = 2 x^2 - [(w+2)/(w+1)] x^4 + {\mathcal O}(x^{5})$. We then conclude that the explicit example presented 
so far fully agrees with the general conclusions obtained in sections \ref{sec2} and \ref{sec3} when the two smearing functions 
are related by the covariant conservation.
 
\subsubsection{Scalar field matter}
\begin{figure}[!ht]
\centering
\includegraphics[height=5cm]{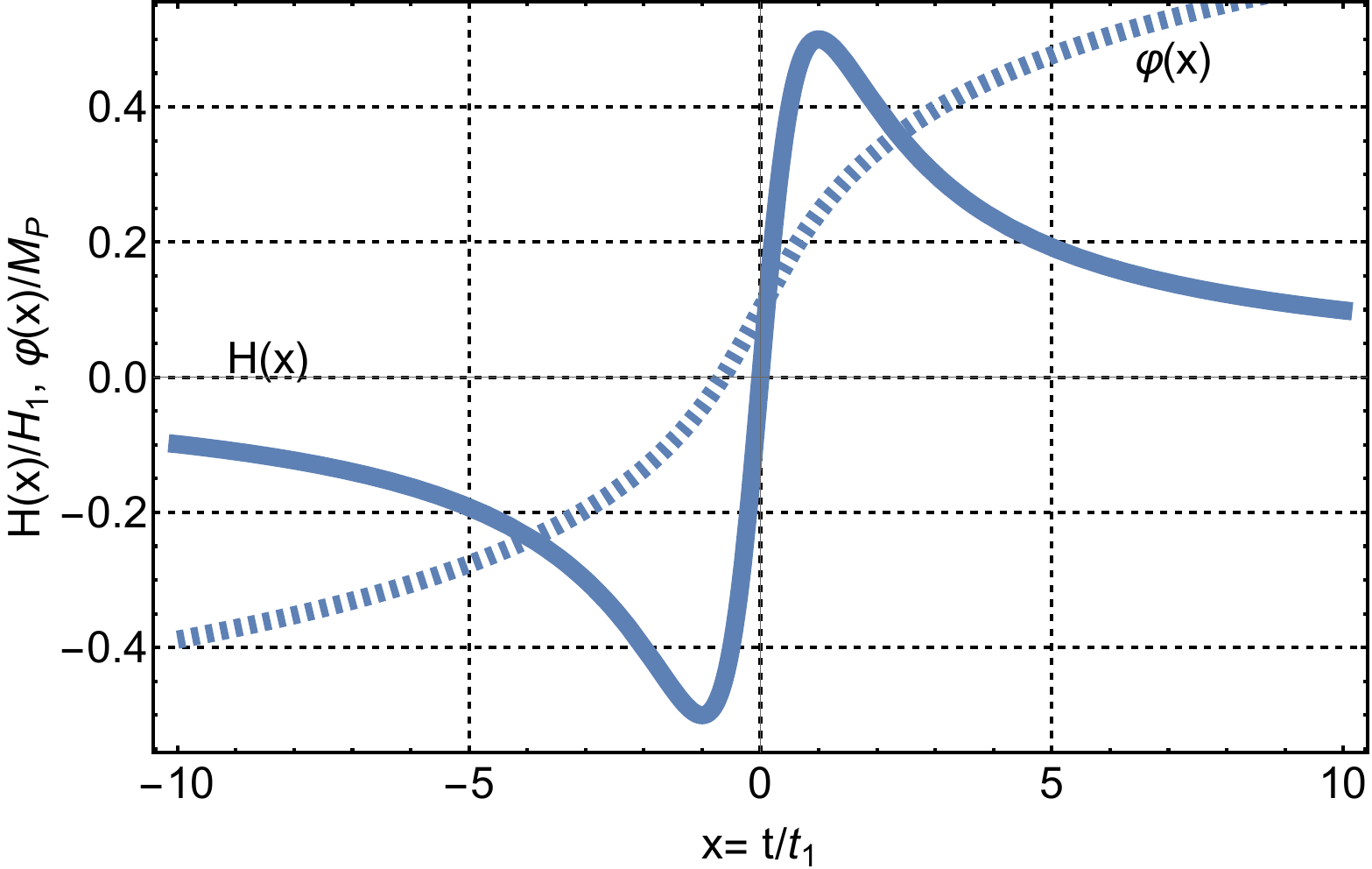}
\includegraphics[height=5cm]{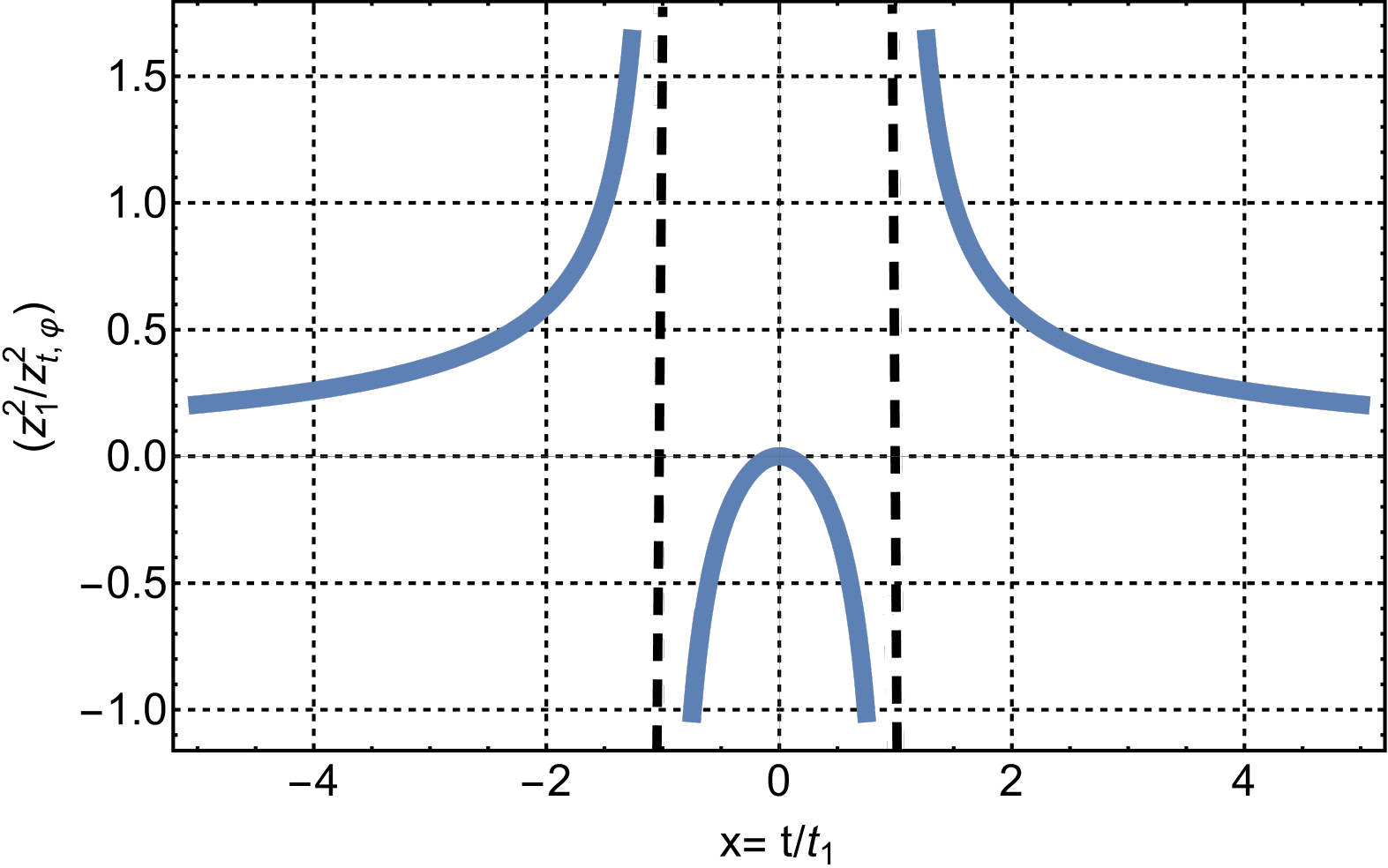}
\caption[a]{In the left plot we illustrate the Hubble rate and the scalar field in the case $V=0$. In the right 
plot we draw instead $\dot{{\mathcal R}}_{k}$ in the large-scale limit; the vertical (dashed) lines in the right plot 
correspond to the singularities of Eq. (\ref{LL18}). Although the underlying background geometry is technically 
regular (see the plot at the left) the evolution of the curvature inhomogeneities is associated to a number 
of pathologies (see plot at the right).}
\label{FIGURE4}      
\end{figure}

In the case of scalar field matter Eqs. (\ref{LL3})-(\ref{LL4}) and (\ref{LL5}) can be solved with the same strategy presented above; a particularly interesting situation is the one where $V =0$ and $\rho_{\varphi}$ reduces to the scalar kinetic term. In this case the evolution of the scalar field follows from the solution of Eq. (\ref{LL5}): 
\begin{equation}
 \varphi(x) = \varphi_{1} \pm \sqrt{\frac{2}{3}} \, \overline{M}_{P} \,\mathrm{arsinh}(x),
\label{LL13}
\end{equation}
while the $\alpha(x)$ and $H(x)$ can be derived from Eqs. (\ref{LL3})--(\ref{LL4}) and are given by 
\begin{equation}
\alpha(x) = (x^2 + 1)^{1/6}, \qquad H(x) = H_{1} \frac{x}{x^2+ 1}, \qquad H_{1} = \frac{1}{3 \, t_{1}}.
\label{LL14}
\end{equation}
Recall, in this respect, that $3 \, H_{1}^2 \, \overline{M}_{P} = \rho_{1}$ and that $ \dot{\varphi}_{1}^2 = 2 \rho_{1}$.
This solution is quantitatively similar to Eq. (\ref{LL8}) in the case $w\to 1$ however the evolution of the curvature 
inhomogeneties needs to be specifically discussed. The first observation is that the instabilities associated 
with the regions where $c_{eff}^2(x)<0$ are not present since the evolution of curvature inhomogeneities
in Fourier space is now given by:
\begin{equation}
\ddot{{\mathcal R}}_{k} + 2 \frac{\dot{z}_{t,\,\varphi}}{z_{t,\,\varphi}} \dot{{\mathcal R}}_{k} + \frac{k^2}{a^2(t)}  {\mathcal R}_{k} =0.
\label{LL15}
\end{equation}
If we go back to Eq. (\ref{HH15}) and 
translate it into the cosmic time parametrization we actually obtain Eq. (\ref{LL15}) (see also appendix \ref{APPB})
where, however, 
\begin{equation} 
z_{t,\varphi}^2 = z_{\varphi}^2 \, a = \frac{a^3 \, \dot{\varphi}^2 \, q}{H^2} = z_{1}^2 \frac{(x^2 -1) \, (x^2 +1)^{1/2}}{x^2},\qquad z_{1}^2= \frac{a_{1}^3 \dot{\varphi}_{1}^2}{H_{1}^2}.
\label{LL16}
\end{equation}
As in the previous case the evolution of the curvature inhomogeneities for $k \ll a H$ follows directly from Eq. (\ref{LL15}) 
and it is given by:
\begin{equation}
{\mathcal R}_{k}(t) = {\mathcal R}_{ex} + \dot{{\mathcal R}}_{ex} \int_{t_{ex}} ^{t} \biggl(\frac{z_{t,\, \varphi}(t_{ex})}{z_{t,\, \varphi}(t^{\prime})}\biggr)^2 \, d\, t^{\prime},
\label{LL17}
\end{equation}
where, by definition, $t_{ex}$ follows from the condition $k = a(t_{ex})\, H(t_{ex})$ so that Eq. (\ref{LL17}) is in fact valid when $ t > t_{ex}$.
Exactly as before we then have that $\dot{\mathcal R}(t) = \dot{{\mathcal R}}_{ex}[z_{t,\, \varphi}(t_{ex})/z_{t,\, \varphi}(t^{\prime})]^2$. This means, once more, that the zeros $z_{t,\varphi}^2$ are the singularities of $\dot{\mathcal R}(t)$ and the divergent contribution explicitly follows from Eq. (\ref{LL16}):
\begin{equation}
\frac{z_{1}^2}{z_{t,\varphi}^2(x)} = \frac{x^2 }{(x^2 -1) \, (x^2 +1)^{1/2}},
 \label{LL18}
 \end{equation}
 implying that $\dot{\mathcal R}(t)$ diverges when $ x \to \pm 1$, i.e. $ t \to \pm t_{1}$. The solution given in Eqs. (\ref{LL13})--(\ref{LL14}) and the condition (\ref{LL18}) are illustrated 
 in Fig. \ref{FIGURE4}.
We can then conclude that the evolution of curvature inhomogeneities gets singular at the onset and at the end of the bouncing regime and this conclusion  matches, incidentally, the one already obtained in Eq. (\ref{LL12}). If more complicated possibilities are considered, the conclusions of Eqs. (\ref{LL12}) and (\ref{LL18}) remain practically unaffected for the class of backgrounds characterized by $\gamma\to 1$ in Eq. (\ref{GG2}). As an example we may discuss the class of models where the potential and the kinetic energy remain in a fixed ratio $\delta$, i.e.  
 $V/\dot{\varphi}^2 = \delta$. In this situation Eq. (\ref{LL5}) can then be solved and the result is: 
\begin{equation}
\ddot{\varphi} = \frac{1}{2 \delta} \biggl(\frac{\partial V}{\partial \varphi}\biggr) \qquad \Rightarrow \qquad \ddot{\varphi} + \frac{3 \, H}{(1 + 2 \delta)} \dot{\varphi} =0,
\label{LL19}
\end{equation}  
 where the first relation follows from the derivation of both sides of $V = \delta\, \dot{\varphi}^2$; the second relation  of Eq. (\ref{LL19}) comes from the first one together with the Klein-Gordon equation of Eq. (\ref{LL5}). The normalized scale 
 factor and energy density become:
 \begin{equation}
 \alpha(x) = (x^2 +1)^{\frac{2 \delta +1}{6}}, \qquad r(x)= \alpha^{-\frac{6}{2 \delta +1}}.
 \label{LL20}
 \end{equation}
 In this case, by going through the same steps outlined before, the divergences in $\dot{{\mathcal R}}$ arise again and generally 
 depend on the values of $\delta$. Various other examples can be discussed along the same lines\footnote{They include, for instance, 
 the cases of oscillating potentials like $V(\varphi) = m^2 \varphi^2/2$ (or, more generally, $V(\varphi) \propto \varphi^{n}$),  
exponential potentials, fast roll potentials characterized by a constant $\ddot{\varphi}/(H\, \dot{\varphi})$.}. In these situations, however, the divergences discussed above always reappear with slightly different features. All in all, the examples discussed in this here corroborate in a specific framework the general conclusions reached in section \ref{sec3}.

We may now get back to the timeline of Fig. \ref{FIGUREDEF1} and note that, with slightly 
qualitative differences, it is also valid in the case of the evolutions described by 
Eqs. (\ref{LL14}) and (\ref{LL20}). For instance in Refs. \cite{DD1,DD2}  Eq. (\ref{LL14}) 
has been used to set the initial conditions of the subsequent inflationary expansion
eventually obtained when $V \gg \dot{\varphi}^2$. The authors actually imagine 
to have $\rho_{\varphi} = \dot{\varphi}^2/2 + V$ and to set the initial data 
when $\dot{\varphi}^2 \gg V$. The solution (\ref{LL14}) is then regarded as an improvement 
in comparison with the standard situation since the initial data can be set already before 
the bounce or, at least, in the regime $0 < t < t_{\ast}$ (see Fig. \ref{FIGUREDEF1}) 
where the evolution corresponds to accelerated expansion (i.e. 
$\dot{a} >0$ and $\ddot{a} >0$). Unfortunately the value of $t_{\ast}$ corresponding 
to Eq. (\ref{LL14}) (and originally introduced in Eq. (\ref{LL8d})) is given by
$t_{\ast} = \sqrt{3/2} \, t_{1}$. But this means that $t_{\ast} > t_{1}$. If the initial data 
would be set prior to $- t_{\ast}$ the curvature inhomogeneities will diverge for $t \to - t_{1}$.
The only consistent choice seems therefore to set the initial conditions 
 for $t\gg t_{1}$ (i.e. in a stage of decelerated expansion) where the curvature inhomogeneities do not diverge.

 \subsection{Bounces of the curvature}
 \subsubsection{Perfect irrotational fluids}
\begin{figure}[!ht]
\centering
\includegraphics[height=5.2cm]{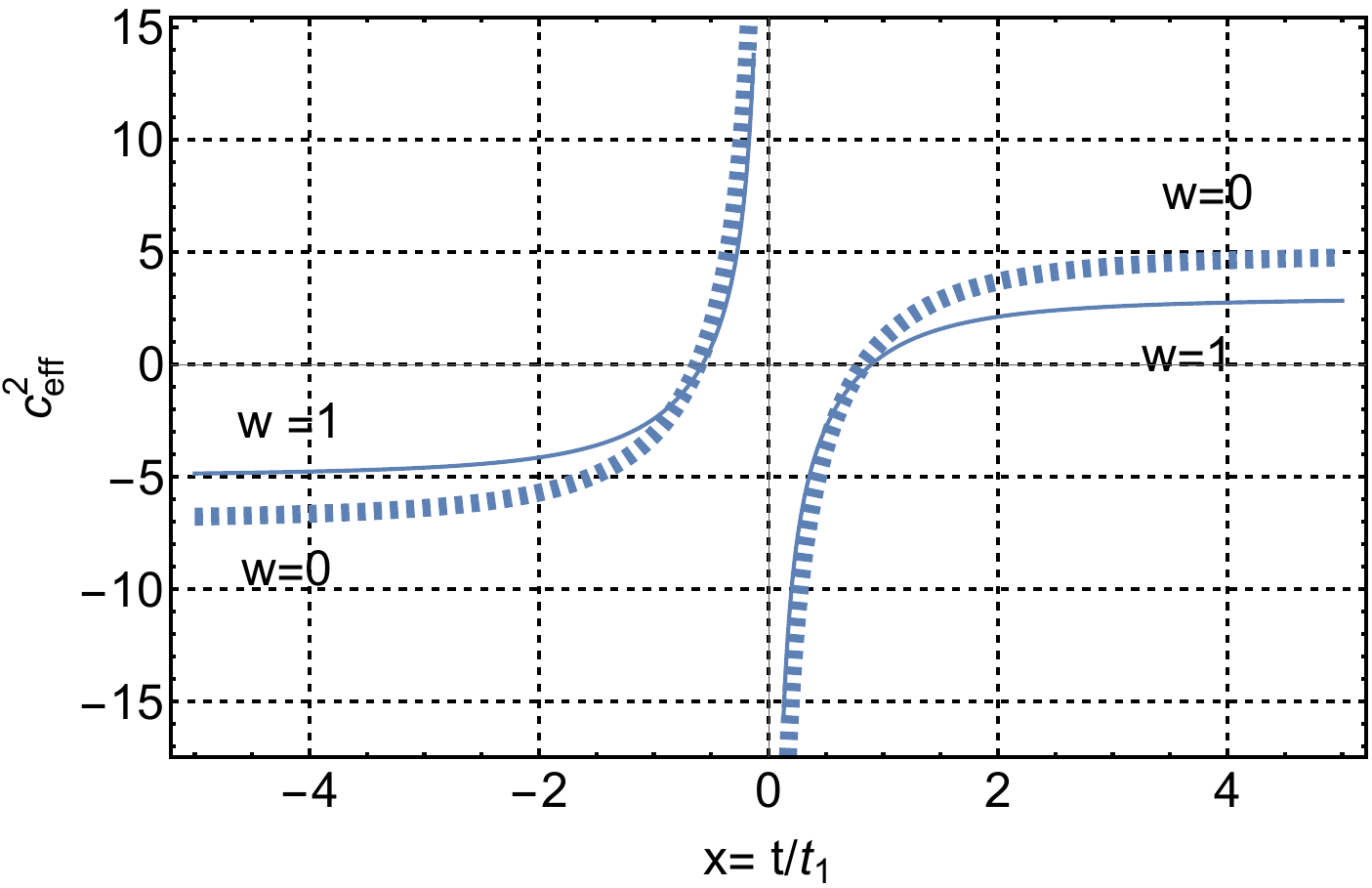}
\includegraphics[height=5.2cm]{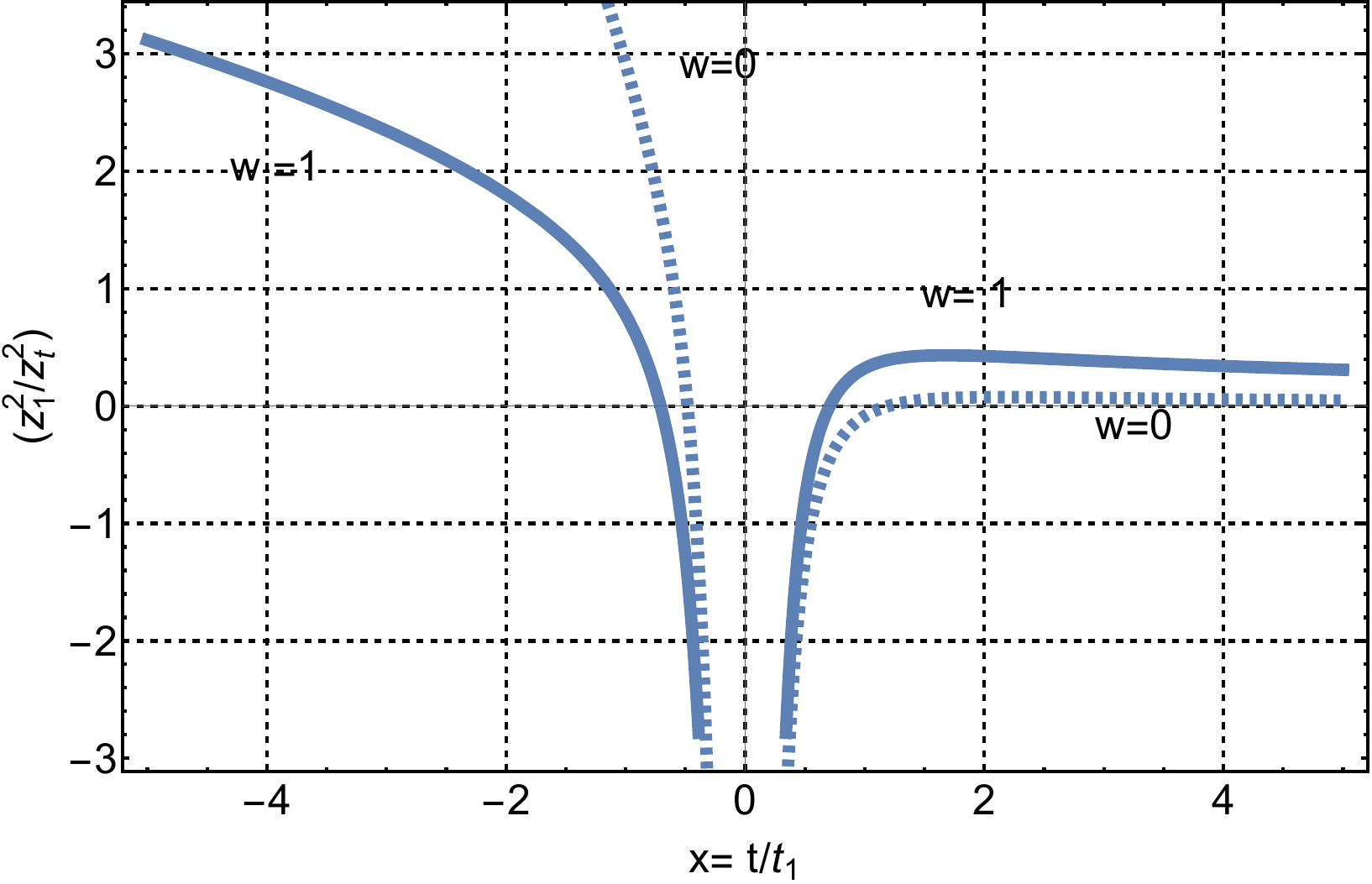}
\caption[a]{In the left plot we illustrate $c_{eff}^2(x)$ for two different values of $w$. In the right plot we instead 
report the evolution of $\dot{{\mathcal R}}$ in the large-scale limit. Both quantities are clearly diverging in the origin 
even if the underlying background does not lead to singularities in the space-time curvature.}
\label{FIGURE5}      
\end{figure}

So far we dealt with the case of the bounces of the scale factor but 
we may now turn to the curvature bounces where, by definition, $f(\rho) > 0$ while 
$q(\rho)$ is related to $f(\rho)$ via Eq. (\ref{CC2}). One of the simplest examples along this direction 
follows from Eq. (\ref{GG6}) by setting $\gamma \to 1$ and $\delta \to 2$:
 \begin{equation}
f(r) = f_{0} \frac{r}{(r^2 + 1)^2}, \qquad\qquad q(r) = \frac{ 2 f_{0} r (1 - r^2 )}{(1 + r^2)^3},
\label{CB1}
\end{equation}
where, as before, $r = \rho/\rho_{1}$; $f_{0}$ is a numerical factor that can be chosen to rationalize 
the solution. The value of $q(r)$ in Eq. (\ref{CB1}) 
is obtained from $f(r)$ thanks to Eq. (\ref{LL2}); only in this case the covariant conservation implies Eq. (\ref{LL5}). If we now insert Eq. (\ref{CB1}) into Eqs. (\ref{LL1}) we obtain 
\begin{equation}
\alpha(x) = \biggl(x + \sqrt{x^2 +1}\biggr)^{\frac{1}{3(w + 1)}}, \qquad  H(x) = \frac{H_{1}}{\sqrt{x^2+1}},\qquad 
H_{1} = \frac{1}{3(w+1) t_{1}},
\label{CB2}
\end{equation}
where $x = t/t_{1}$ and $3\, H_{1}^2 \, \overline{M}_{P}^2 = \rho_{1}$ provided $f_{0} =4$. Equation (\ref{CB2}) implies, as expected, that $H^2(x)$ does not vanish while $\dot{H}$ vanishes in $t =0$. Since $\dot{a} >0$, the solution (\ref{CB2}) is always expanding however 
$\ddot{a}$ may be either positive or negative.  The effective sound speed governing the evolution of curvature inhomogeneities can be directly obtained from Eq. (\ref{FF16}):
\begin{equation}
c_{eff}^2(r) = c_{s,\,t}^2 + (w + 1) \frac{3 r^4 - 8 r^2 +1}{1 - r^4}.
\label{CB3}
\end{equation}
In spite of the positivity of $f(r)$ the effective sound speed appearing in the evolution 
of the curvature inhomogeneities diverges as a function of $x$; indeed 
from Eq. (\ref{CB3}) we get
\begin{equation}
c_{eff}^2(x) = c_{s,\,t}^2 + ( 1 + w) \biggl[ \frac{2 x^2 - 1}{x \sqrt{x^2 +1}} -1 \biggr],
\label{CB4}
\end{equation}
that is singular in the origin. In the left plot of Fig. \ref{FIGURE5} we illustrate $c_{eff}^2$.
As for the bounces of the scale factor, $c_{eff}^2 <0$ in the asymptotic region before 
the bounce; this makes the evolution of the curvature inhomogeneities particularly unrealistic.
Indeed that the evolution of ${\mathcal R}$ 
is unstable throughout the whole pre-bouncing phase and even in the decelerated epoch at late times. We can finally compute $z_{t}^2(x)$ which controls the divergences 
of $\dot{\mathcal R}$ in the large-scale limit:
\begin{equation}
z^2(x) = 2\,z_{1}^2  \frac{ (1+w)\, x^2 \, \bigl( x + \sqrt{x^2 +1}\bigr)^{\frac{1}{w+1}}}{(w -1) \, x\,\sqrt{x^2 +1} + (w+1) ( 2 x^2 -1)}, \qquad z_{1}^2 = \frac{a_{1}^3 \rho_{1}}{H_{1}^2}.
\label{CB5}
\end{equation}
Once more, the zeros of Eq. (\ref{CB5}) imply a divergence in $\dot{{\mathcal R}}$ and in the right plot of Fig. \ref{FIGURE5}
we indeed illustrated $(z_{1}/z_{t})^2$ for two different values of $w$; we then see that $\dot{\mathcal R}$ 
 diverges, as expected, in the large-scale limit. All in all we can say that the small-scale gradient instabilities 
 and the large-scale divergences make this evolution unrealistic even though the underlying 
 background evolution does not diverge.

\subsubsection{Scalar field matter}
We can now swiftly investigate the curvature bounces obtained for scalar 
field matter always in the case $\gamma \to 1$ and $\delta \to 2$ (see Eq. (\ref{GG6}) and discussion therein); the difference is that now the two smearing functions depend on $\rho_{\varphi}$ and not on the fluid energy density. As expected the gradient instability associated with the regions $c_{eff}^2(x)<0$ does not 
appear in the scalar field case since the relevant evolution equation does not involve any $c_{eff}^2$. 
According to the general considerations of Eqs. (\ref{HH20})--(\ref{HH21}) 
$\dot{{\mathcal R}}$ should not diverge when $(\gamma - 2 \delta +1) >0$. 
However, since in this case we explicitly assume $\gamma \to 1$ and $\delta \to 2$ we expect that 
the curvature inhomogeneities will still be divergent. Since the presence of the potential does not 
crucially modify the conclusions in the regime where $\dot{\varphi}^2$ dominates, we consider  the case where the potential vanishes. If $V =0$ we then have that the solution of Eq. (\ref{LL3}) is:
\begin{equation}
\alpha(x) = \biggl( x + \sqrt{x^2 +1}\biggr)^{1/6}, \qquad r(x) = \alpha^{-6},
\label{CB6}
\end{equation}
where we used the profile of Eq. (\ref{CB1}) with $r= \rho_{\varphi}/\rho_{1}$.
We can therefore deduce $z_{t,\,\varphi}^2$ which is given by:
\begin{equation}
\frac{z_{t,\, \varphi}^2(x)}{z_{1}^2} =   \frac{2 \, x \, \sqrt{x + \sqrt{x^2 +1}}}{\sqrt{x^2 +1}}.
\label{CB7}
\end{equation}
Since $z_{t,\, \varphi}^2(x)$ vanishes for $x\to 0$, as expected, its inverse diverges.
\begin{figure}[!ht]
\centering
\includegraphics[height=5.2cm]{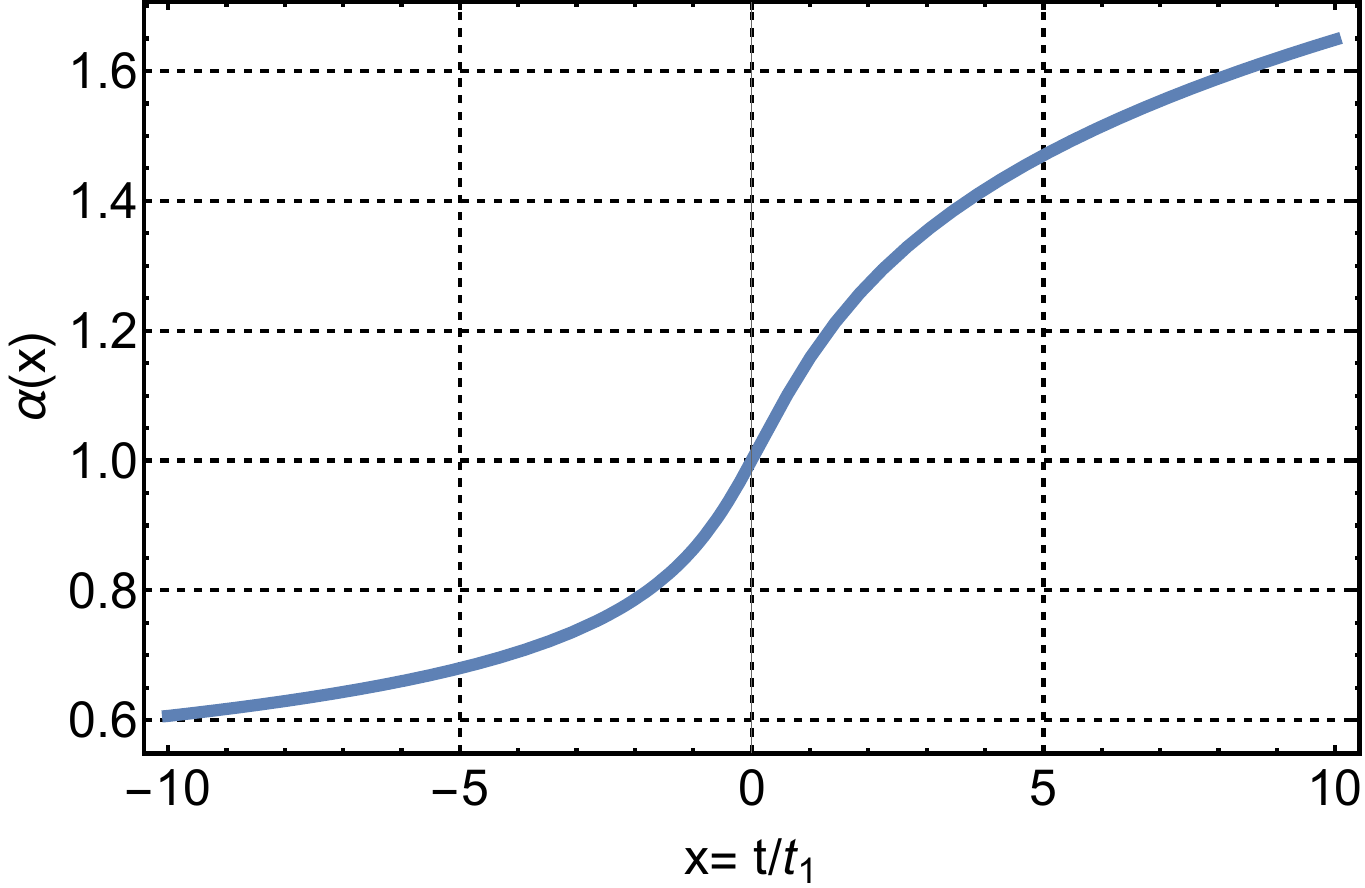}
\includegraphics[height=5.2cm]{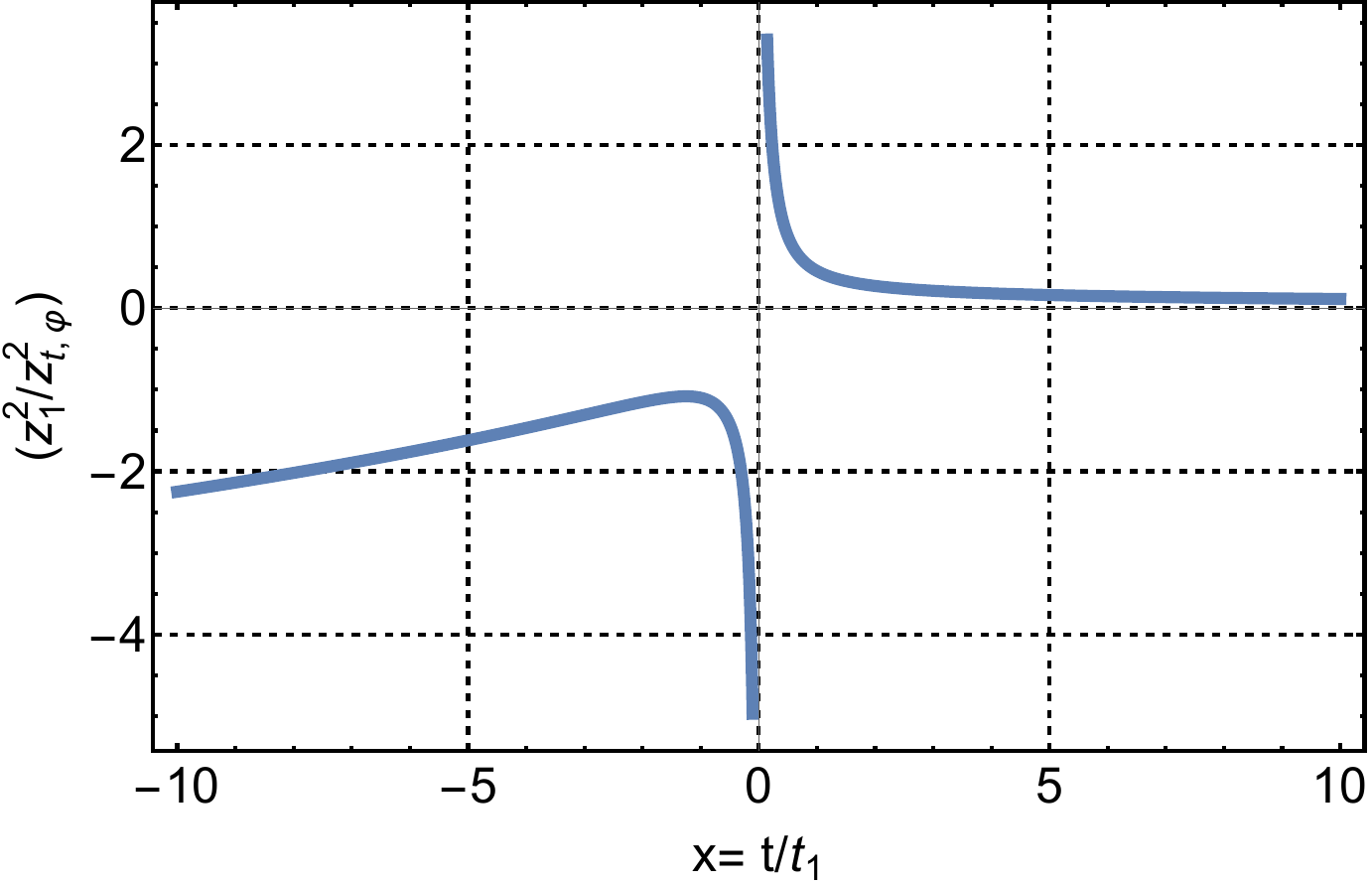}
\caption[a]{In the left plot we illustrate the normalized scale factor while in the right plot 
we draw the leading contribution to $\dot{{\mathcal R}}$. The divergence in the right 
plot corresponds, as expected, to the zeros of $z_{t,\, \varphi}^2(x)$. Again 
the divergence in $(z_{1}^2/z_{t,\,\varphi}^2)$ arises even if the underlying background 
geometry is technically regular.}
\label{FIGURE6}      
\end{figure}
This conclusion is also illustrated in Fig. \ref{FIGURE6} where we clearly see that 
$z_{1}^2/z_{t, \,\varphi}^2$ diverges. In the light of Eqs. (\ref{HH20})--(\ref{HH21}) this 
conclusion is not surprising since, as stressed above, when
 $\gamma \to 1$ and $\delta \to 2$ the divergences are expected. According to Eq. (\ref{HH21}) the regular evolution 
of curvature inhomogeneities is expected for $(\gamma - 2 \delta +1) >0$. In this portion of the 
parameter space it is however difficult to find analytic solutions expressible in a closed form. We 
leave this possibility for future analysis but stress here that the explicit solutions of this section fully corroborate the general conclusions of section \ref{sec3}.

\subsection{The fluctuations of the density contrast}
In the previous considerations we explicitly analyzed the evolution of ${\mathcal R}$ and of its time derivative 
in different backgrounds. We now observe that the evolution of ${\mathcal R}$ and ${\mathcal R}^{\prime}$ 
 directly determines the evolution of the density contrast. In appendix \ref{APPC} we introduced 
the gauge-invariant density contrasts $\zeta$ and $\zeta_{\varphi}$ obeying Eq. (\ref{APPC5}) which we rewrite here 
for the sake of convenience:
 \begin{equation}
 \zeta = {\mathcal R} - \frac{\dot{{\mathcal R}}}{3 \, H \, c_{eff}^2(\rho)}, \qquad  \zeta_{\varphi} = {\mathcal R} - \frac{\dot{{\mathcal R}}}{3\, H}.
 \label{MDENS1}
 \end{equation}
Thanks to the analysis of appendix \ref{APPC} we can appreciate that, in the gauge of Eq. (\ref{FF1}), $\zeta$ and $\zeta_{\varphi}$ coincide with the density contrasts 
so that 
\begin{equation}
\zeta = \frac{\delta_{s} \rho}{3 ( \rho + p)}, \qquad \zeta_{\varphi} =  \frac{\delta_{s} \rho_{\varphi}}{3 ( \rho_{\varphi} + p_{\varphi})}.
\label{MDENS2}
\end{equation}
Since $\zeta$ and $\zeta_{\varphi}$ equal the density contrasts on the hypersurfaces where the spatial 
curvature is uniform (as implied by the gauge (\ref{FF1})), the gauge-invariant relation (\ref{MDENS1}) implies
that $\zeta$ and $\zeta_{\varphi}$ (and the corresponding density contrasts) diverge at the same points 
where ${\mathcal R}$ and $\dot{{\mathcal R}}/H$ are singular. 
In the case of curvature bounces $H$ does not vanish; therefore in this case it follows that the divergences of $\zeta$ and 
$\zeta_{\varphi}$ coincide exactly with the ones of ${\mathcal R}$ and $\dot{{\mathcal R}}$. 
For the bounces of the scale factor we have that $H\to 0$ at the end of the contracting stage. In this case it may happen 
that the divergences of $1/H$ may compensate the ones of $\dot{{\mathcal R}}$.  This is, unfortunately, not what 
happens for the explicit examples considered before in this section. It then follows 
that the divergences appearing in $\dot{{\mathcal R}}$ can be viewed as singularities 
of the density contrast unless, in the specific background, ${\mathcal R}$ and $\dot{{\mathcal R}}/H$ are both finite.
The results of Eqs. (\ref{MDENS1})--(\ref{MDENS2}) demonstrate once more that while 
the fuzzy bounces described in this paper might seem technically regular, the evolutions of the density contrast and of the 
curvature inhomogeneities restrict the range of validity of the solutions derived with this method.

Despite the singularities associated with the evolution of the curvature inhomogeneities it would be 
tempting to derive the large-scale spectral of ${\mathcal R}$. This exercise seems a bit sterile as long as the 
regularity of the background solution is not 
associated with a non-singular evolution of ${\mathcal R}$ and $\zeta$. We may however pretend that 
the evolution before the bounce is complemented at late time by a radiation-dominated stage. In this 
case the spectral slope of $k^{3} \,|{\mathcal R}_{k}|^2$ could simply be deduced by approximately 
solving the evolution of ${\mathcal R}$ in the region that precedes the bounce. Let us then focus, as 
an example, on the case of the bounces of the scale factor and let us 
specifically consider the example of Eq. (\ref{LL8}). In this case the background contracts, at early time, 
as $a(t)= (- t/t_{1})^{2/[3(w+1)]}$. From the other features of the solution we can solve Eq. (\ref{APPB1}) 
for $t \ll -t_{1}$ and pretend (indeed rather baldly) that the evolution of the curvature inhomogeneities 
will be non-singular and well behaved in a putative radiation-dominated stage taking place at late times.
In this situation the power spectrum of curvature inhomogeneities 
is characterized by a slope $n_{{\mathcal R}}$:
\begin{equation}
P_{\mathcal R}(k,\tau) = \frac{k^3}{2 \pi^2} \bigl|{\mathcal R}_{k}(\tau)\bigr|^2 \propto \biggl(\frac{k}{a \, H}\biggr)^{n_{{\mathcal R}}}, \qquad n_{{\mathcal R}} = 3 - \frac{3( 1 - w)}{3 w +1}.
\label{MDENS3}
\end{equation}
Concerning this result few comments are in order.
\begin{itemize}
\item{} Equation (\ref{MDENS3}) hold in the case of a perfect fluid background but a similar result holds in the case of scalar field 
matter; for instance for $w \to 1$ (corresponding to the case of a dominant kinetic term of the scalar field $\varphi$) the 
spectral slope becomes $n_{{\mathcal R}} \to 3$;
\item{} the amplitude of the power spectrum delicately depends on the intermediate regime and it makes therefore 
no sense to predict it unless a non-singular evolution of the curvature inhomogeneities 
is reliably deduced;
\item{} a quasi flat spectrum of curvature inhomogeneities is realized when $w\to 0$ even if, as already mentioned, 
this interesting possibility is hard to justify in the light of the caveats propounded in the present and in the previous 
section.
\end{itemize}
To conclude this discussion we can finally consider the $n_{{\mathcal R}}$ obtainable in the case 
of the curvature bounces of Eq. (\ref{CB2}) and within the same (partially unrealistic) hypotheses 
leading to Eq. (\ref{MDENS3}). By repeating the same calculation in the case of Eq. (\ref{CB2}) we would have 
that 
\begin{equation}
n_{\mathcal R} = \frac{6(w+1)}{3 w + 4},
\label{MDENS4}
\end{equation}
and the most notable aspect of this result seems to be the absence, even in principle, of a quasi-flat spectrum.

\subsection{Some related remarks}
Thanks to the contracted Bianchi identities and to the covariant conservation of the modified energy-momentum tensor, the relation between $f(\rho)$ and $q(\rho)$ (or between $f(\rho_{\varphi})$ and $q(\rho_{\varphi})$) is not arbitrary. Furthermore, when $f(\rho)\to 0$ (or $f(\rho_{\varphi}) \to 0$) we also have that $H \to 0$ and this situation corresponds, broadly speaking, to a bounce of the scale factor where a phase of accelerated contraction turns into a stage of accelerated expansion. In this case the evolution of curvature inhomogeneities is
 characterized by a gradient instability since the square of the effective sound speed 
 may become negative;  the time derivative of the curvature inhomogeneities on comoving orthogonal hypersurfaces is also potentially divergent in the large-scale limit. If the energy-momentum tensor is modelled in terms of a scalar field,
 the gradient instabilities do not appear while the divergences in the curvature 
 inhomogeneities remain unaffected. The toy models associated with $f(\rho)\to 0$ and $f(\rho_{\varphi})\to 0$ are potentially interesting but fail to address the problem of the singularity. The bounces of the curvature (corresponding to $f(\rho)>0$ and $f(\rho_{\varphi}) > 0$) imply instead a milder degree of divergence of the curvature inhomogeneities so that the underlying solutions seem preferable for the purpose of setting the initial conditions of the subsequent 
 inflationary expansion.  We finally mention that the discussion of this section focussed
 primarily on the case of the adiabatic fluctuations corresponding to ${\mathcal S}_{{\mathcal R}}\to 0$ (see Eqs. (\ref{FF14})--(\ref{FF17}) and discussions thereafter). While this choice 
 is motivated by the relevance of the adiabatic fluctuations in the current paradigm of structure 
 formation \cite{ADP1,ADP2} the results of sections \ref{sec3} and \ref{sec4} can be easily 
 generalized to the situation where the non-adiabatic (or entropic) fluctuations 
 are present. The differences between
adiabatic and non-adiabatic fluctuations are discussed, for instance, in Refs. \cite{WAD1} (see also \cite{nad3}). The relevance of non-adiabatic fluctuations for the initial data of the cosmic microwave 
background anisotropies has been discussed along various perspectives (see, for instance, Refs. \cite{nad1,nad1a,nad2,nad2a}). The inclusion of the entropic fluctuations will not be discussed 
specifically in this paper even if a preliminary analysis suggests the same drawbacks 
already pointed out in the adiabatic case.

\newpage
\renewcommand{\theequation}{5.\arabic{equation}}
\setcounter{equation}{0}
\section{Concluding considerations}
\label{sec5}
The energy and the enthalpy densities can 
be smeared by a pair of independent form factors that are connected by the contracted Bianchi identities and, in this case, the evolution both of the background and of the curvature inhomogeneities is determined by the properties of the smearing functions. A bounce of the scale factor corresponds to a vanishing Hubble rate in the neighbourhood of the maximal curvature scale and in this situation a stage of accelerated contraction turns into a phase of decelerated expansion even if the intermediate region may be characterized by a short epoch of accelerated expansion. This intermediate epoch cannot be used, however, to set the initial data of the conventional inflationary phase since the curvature inhomogeneities diverge exactly in the same range of the dynamical evolution.  In the complementary perspective of a curvature bounce, the time derivative of the extrinsic curvature vanishes while the Hubble rate itself does not have any zero. For an irrotational fluid the evolution of curvature inhomogeneities always inherits an effective sound speed that depends on the smearing functions and on their derivatives with respect to the energy density. The square of the effective sound speed may get negative both for the bounces of the scale factor and of the curvature: this occurrence signals the presence of fatal instabilities that disappear in the case of scalar field matter. In the latter case, however, the derivative of curvature inhomogeneities is generally divergent across the bounce.

The obtained results have been corroborated by specific examples that confirmed the background-independent expectations.  The bouncing models constructed in this manner could be then viewed as a possible alternative to inflation or just as a completion of a subsequent stage of accelerated expansion. In the first option the spatial curvature and the initial inhomogeneities could be suppressed during the bouncing regime. However, even if this happens, the evolution of the corresponding curvature inhomogeneities is always singular. This phenomenon suggests that the singularity in the curvature is eliminated from the background but it reappears in the corresponding fluctuations. If the present construction is viewed as complementary to inflation, then the initial data of the inflaton cannot be assigned before a bounce of the scale factor but they can be probably assigned prior to a curvature bounce. 

\section*{Acknowledgements}
The author wishes to thank A. Gentil-Beccot, L. Pieper and S. Rohr of the CERN scientific 
information service for their kind help. 

\newpage

\begin{appendix}
\renewcommand{\theequation}{A.\arabic{equation}}
\setcounter{equation}{0}
\section{Complements on the curvature inhomogeneities}
\label{APPA}
Some general results on the scalar fluctuations in the gauge defined by Eq. (\ref{FF1}) will now 
be recalled. While the use of this gauge is not mandatory 
(indeed more standard gauges can also be employed) it turns out, however, 
that the coordinate system of Eq. (\ref{FF1}) simplifies the derivation of the evolution equation of the curvature inhomogeneities for the specific class of problems discussed in this paper. 
With these necessary caveats, in the gauge (\ref{FF1}), the scalar 
fluctuations of the Einstein tensor become  
\begin{eqnarray}
\delta_{s} {\mathcal G}_{0}^{\,\,0} &=&  \frac{2}{a^2} \biggl[- {\mathcal H} \nabla^2 B - 3 {\mathcal H}^2 \phi \biggr],
\label{APPA1}\\  
\delta_{s} {\mathcal G}_{i}^{j} &=& \frac{1}{a^2} \biggl\{ \bigg[ - 
2 ({\mathcal H}^2 + 2 {\mathcal H}') \phi - 2 {\mathcal H} \phi' \biggr]
- \nabla^2 (  B^{\prime} + 2 {\mathcal H} B + \phi)\biggr\} \delta_{i}^{\,\,j}
\nonumber\\
&+& \frac{1}{a^2}\partial_{i}\partial^{j} \biggl[  B^{\prime} + 2 {\mathcal H} B + \phi \biggr],
\label{APPA2}\\
\delta_{s} {\mathcal G}_{i}^{0} &=& \frac{2 {\mathcal H}}{a^2} \partial_{i} \phi.
\label{APPA3}
\end{eqnarray} 
In the gauge (\ref{FF1}) the  Bardeen's potentials correspond, respectively, to 
\begin{equation}
\Phi = \phi + {\mathcal H} B + B^{\prime}, \qquad \qquad \Psi = - {\mathcal H} B,
\label{APPA4}
\end{equation}
where, as usual, $a(\tau)$ is the scale factor ${\mathcal H} = a^{\prime}/a$ and the 
prime denotes a derivation with respect to the conformal time coordinate $\tau$.
We stress that both $\Phi$ and $\Psi$ are invariant under infinitesimal coordinate 
transformations even if their expression can be evaluated (and employed) in any gauge.
In the gauge (\ref{FF1}) the difference between the Bardeen's potentials (usually related to the anisotropic stress) reproduces the combination appearing in Eq. (\ref{APPA2})
\begin{equation}
\Phi - \Psi = \phi + B^{\prime} + 2 {\mathcal H} B.
\label{APPA5}
\end{equation}
In the same spirit the scalar fluctuations of the spatial curvature on comoving orthogonal hypersurfaces are gauge-invariant and can be expressed as 
\begin{equation}
\delta_{s} \,\,^{(3)}R = - \frac{4}{a^2} \nabla^2 {\mathcal R}.
\label{APPA6}
\end{equation}
Again ${\mathcal R}$ is invariant under infinitesimal coordinate transformations
but in the gauge of Eq. (\ref{FF1}) it assumes a particularly simple form, namely 
\begin{equation}
{\mathcal R} = - \frac{{\mathcal H}^2}{{\mathcal H}^2 - {\mathcal H}^{\prime}} \phi.
\label{APPA7}
\end{equation}
Finally since ${\mathcal R}$ is gauge-invariant it can be directly expressed in terms of the Bardeen's potentials $\Phi$ and $\Psi$:
\begin{equation}
{\mathcal R} = - \Psi - \frac{{\mathcal H}({\mathcal H} \Phi + \Psi^{\prime})}{{\mathcal H}^2 - {\mathcal H}^{\prime}}.
\label{APPA8}
\end{equation}
If we now insert Eq. (\ref{APPA4}) in Eq. (\ref{APPA8}) we can again obtain Eq. (\ref{APPA7}).
\renewcommand{\theequation}{B.\arabic{equation}}
\setcounter{equation}{0}
\section{Transformations of the canonical pump fields}
\label{APPB}
In the bulk of the paper the conformal and cosmic time 
parametrizations have been employed interchangeably since 
their relation stipulates that $a(\tau) \, d\tau = d\,t$. The prime always denotes a derivation with respect to $\tau$ while the overdots indicate 
the derivations with respect to the cosmic time $t$ and the transition from one description 
to the other is trivial for many related quantities. For instance it is well known 
that ${\mathcal H} = a \, H$ or that $({\mathcal H}^2 - {\mathcal H}^{\prime}) = - a^2 \dot{H}$ and so on and so forth. In the case of the evolution of the curvature inhomogeneities (and more generally for any second-order differential equation) the connection between the two parametrizations involves a formal modification of the pump fields. To avoid potential confusions and misunderstandings we therefore recall that if we start from the evolution of the curvature inhomogeneities written in conformal time 
coordinate
\begin{equation}
{\mathcal R}_{k}^{\prime\prime} + 2 \frac{z^{\prime}}{z} {\mathcal R}_{k}^{\prime} + k^2 \, c_{eff}^2 {\mathcal R}_{k} =0,
\label{APPB1}
\end{equation}
the cosmic time counterpart of Eq. (\ref{APPB1}) is obtained after some simple algebra 
and it is given by:
\begin{equation}
\ddot{{\mathcal R}}_{k}+ \biggl(2 \frac{\dot{z}}{z} + H \biggr)\dot{{\mathcal R}}_{k} + k^2 \, \frac{c_{eff}^2}{a^2} {\mathcal R}_{k} =0.
\label{APPB2}
\end{equation}
We may now introduce a new pump field $z_{t}$ which is actually defined in the following manner:
\begin{equation}
z_{t}^2(t) = z^2(t) \, a(t) = \frac{a^{5} (\rho + p)\, q(\rho)}{{\mathcal H}^2 \,\, c_{eff}^2} \equiv \frac{a^3 (\rho+ p)\, q(\rho)}{ H^2 \, c_{eff}^2}.
\label{APPB3}
\end{equation}
The result of Eq. (\ref{APPB3}) implies that Eq. (\ref{APPB2}) can also be written as: 
\begin{equation}
\ddot{{\mathcal R}}_{k}+ 2 \frac{\dot{z_{t}}}{z_{t}} \dot{{\mathcal R}}_{k} +  \frac{k^2 \, c_{eff}^2}{a^2} {\mathcal R}_{k} =0.
\label{APPB4}
\end{equation}
By the same token, in the scalar field case, we shall be able to introduce a rescaled variable (be it $z_{\varphi,\,t}$) whose explicit relation with $z_{\varphi}$ is simply given by:
\begin{equation}
z_{\varphi,\, t}^2 = z_{\varphi}^2 \, a= \frac{a^3 \, \varphi^{\prime\, 2}\, q(\rho_{\varphi}) }{{\mathcal H}^2 } = \frac{ a^{3} \, \dot{\varphi}^2\, q(\rho_{\varphi})}{H^2}.
\label{APPB5}
\end{equation}
The analog of Eq. (\ref{APPB4}) in the scalar field case becomes therefore: 
\begin{equation}
\ddot{{\mathcal R}}_{k}+ 2 \frac{\dot{z_{\varphi,\,t}}}{z_{\varphi,\,t}} \dot{{\mathcal R}}_{k} +  \frac{k^2}{a^2} {\mathcal R}_{k} =0.
\label{APPB6}
\end{equation}
The results of Eqs. (\ref{APPB4}) and (\ref{APPB6}) are relevant in the analyses of section 
\ref{sec4} where the general theoretical expectations (based on background-independent 
considerations) have been corroborated by a number of explicit examples.
\renewcommand{\theequation}{C.\arabic{equation}}
\setcounter{equation}{0}
\section{The evolution of the density contrast}
\label{APPC}
It is generally true that when ${\mathcal R}$ and ${\mathcal R}^{\prime}$ are known, the evolution 
of the density contrast follows easily. For this purpose it is useful to remark that the density 
contrast has a gauge-invariant meaning conventionally denoted by  $\zeta$:
\begin{equation}
\zeta = - \biggl(\Psi + \frac{{\mathcal H} \delta^{(\mathrm{gi})}\rho}{\rho^{\prime}}\biggr), \qquad \delta^{(\mathrm{gi})}\rho = \delta_{s} \rho + \rho^{\prime} \, B,
\label{APPC1}
\end{equation}
where $\Psi$ is the Bardeen potential already introduced in Eq. (\ref{APPA4}) and $\delta^{(\mathrm{gi})}\rho$ is the gauge-invariant 
fluctuation of the energy density coinciding with $(\delta_{s} \rho + \rho^{\prime} \, B)$ in the gauge of Eq. (\ref{FF1}) where it is also true that
$\Psi = - {\mathcal H} B$. It then follows that in the gauge of Eq. (\ref{FF1}) $\zeta$ is nothing but the density contrast:
\begin{equation}
\zeta = - \frac{{\mathcal H} \delta_{s} \rho}{\rho^{\prime}} = \frac{ \delta_{s} \rho}{3 \, ( \rho +p)},
\label{APPC2}
\end{equation}
where we implicitly used the result of Eq. (\ref{CC2}) imposing a specific relation between $q(\rho)$ and $f(\rho)$. The same result of Eq. (\ref{APPC2}) 
holds in the case of scalar field matter so that $\zeta_{\varphi} = \delta_{s} \rho_{\varphi}/[3 ( \rho_{\varphi} + p_{\varphi})]$. The relation between $\zeta$, ${\mathcal R}$ and $\Psi$ can be written in gauge-invariant terms and it is given by:
\begin{equation}
\zeta = {\mathcal R} + \frac{2\,\nabla^2 \Psi}{3\,\ell_{P}^2 \,a^2 \, q(\rho) (\rho+ p)}, \qquad \zeta_{\varphi} = {\mathcal R} + \frac{2\,\nabla^2 \Psi}{3\, \ell_{P}^2 \,q(\rho_{\varphi}) \varphi^{\prime\, 2}}.
\label{APPC3}
\end{equation}
It is not difficult to see that Eq. (\ref{APPC3}) has exactly the same content of the Hamiltonian constraint; indeed Eq. (\ref{FF9}) coincides with Eq. (\ref{APPC3}) by recalling that, in the gauge (\ref{FF1}), 
 $\Psi = - {\mathcal H} B$, ${\mathcal R}= - \phi\, ({\mathcal H}^2 - {\mathcal H}^{\prime})/{\mathcal H}^2$ and $ \delta_{s} \rho = 3 (\rho + p) \zeta$. 
 The same conclusion follows in the case of scalar field matter. In the absence of non-adiabatic pressure fluctuations 
 Eqs. (\ref{FF18}) and (\ref{FF20}) relate $\Psi$ (or $B$) to ${\mathcal R}^{\prime}$. It then follows immediately
 that $\zeta$ and $\zeta_{\varphi}$ can be written, respectively as 
 \begin{equation}
 \zeta = {\mathcal R} - \frac{{\mathcal R}^{\prime}}{3 {\mathcal H} \, c_{eff}^2(\rho)}, \qquad  \zeta_{\varphi} = {\mathcal R} - \frac{{\mathcal R}^{\prime}}{3 {\mathcal H}}.
 \label{APPC4}
 \end{equation}
 In terms of the cosmic time parametrization Eq. (\ref{APPC4}) becomes:
   \begin{equation}
 \zeta = {\mathcal R} - \frac{\dot{{\mathcal R}}}{3 \, H \, c_{eff}^2(\rho)}, \qquad  \zeta_{\varphi} = {\mathcal R} - \frac{\dot{{\mathcal R}}}{3\, H}.
 \label{APPC5}
 \end{equation}
 Equations (\ref{APPC4})--(\ref{APPC5}) have a gauge-invariant 
meaning and have been derived without imposing any large-scale limit; this means that they are valid both at small and large-scales. It is then clear from Eqs. (\ref{APPC4})--(\ref{APPC5}) that the density contrast inherits the divergences of ${\mathcal R}$ 
and ${\mathcal R}^{\prime}$ (or $\dot{\mathcal R}$). It is then sufficient to solve the evolution of ${\mathcal R}$ and ${\mathcal R}^{\prime}$ (or $\dot{{\mathcal R}}$)
in the large-scale limit (as discussed in sections \ref{sec3} and \ref{sec4}) to deduce the singularities of the density contrast.
\end{appendix}


\newpage
\begin{thebibliography}{99}

\itemsep -3pt

\bibitem{WW1} D.~N.~Spergel {\it et al.},   Astrophys.\ J.\ Suppl.\  {\bf 148}, 175 (2003).

\bibitem{WW2} D.~N.~Spergel {\it et al.}, Astrophys.\ J.\ Suppl.\ \ {\bf 170}, 377 (2007).

\bibitem{WW2a} L.~Page {\it et al.}, Astrophys.\ J.\ Suppl.\  {\bf 170}, 335 (2007).

\bibitem{WW2b} C.~L.~Bennett {\it et al.}, Astrophys.\ J.\ Suppl.\    {\bf 208} 20 (2013).

\bibitem{WW3}  N.~Aghanim {\it et al.} [Planck Collaboration], Astron. Astrophys. {\bf 641}, A6 (2020).

\bibitem{WW4} P.~A.~R.~Ade {\it et al.} [BICEP and Keck], Phys. Rev. Lett. {\bf 127}, 151301 (2021).

\bibitem{WW5} A.~A.~Starobinsky, Phys. Lett. B \textbf{91}, 99 (1980).

\bibitem{WW5a} A.H. Guth, Phys. Rev. D {\bf 23}  347 (1981).

\bibitem{WW5b} A.D. Linde, Phys. Lett. B {\bf 108}, 389 (1982).

\bibitem{WW5c}  A. Albrecht, P.J. Steinhardt, Phys. Rev. Lett. {\bf 48}, 1220 (1982).

\bibitem{WW6} S. Weinberg, {\it Cosmology} (Oxford University Press, Oxford, UK, 2008).

\bibitem{ADP1} P. J. E.  Peebles, {\it Principles of Physical Cosmology},  (Princeton
University Press, Princeton 1993).

\bibitem{ADP2} P. J. E. Peebles  and J. T. Yu, Astrophys. J. {\bf 162},  815 (1970).

\bibitem{WW7} P. J. Steinhardt,  Sci. Am. {\bf 304}, 18 (2011)

\bibitem{WW7a} A. Ijjasa, P.  Steinhardt,  A. Loeb, Phys. Lett.  B {\bf 723}, 261 (2013).

\bibitem{WW8} D.~S.~Salopek and J.~M.~Stewart, Class.\ Quant.\ Grav.\  {\bf 9}, 1943 (1992).

\bibitem{WW8a} N. Deruelle and K. Tomita, Phys. Rev. D {\bf 50}, 7216 (1994).

\bibitem{WW8b}  M.~Giovannini, Phys.\ Lett.\ B {\bf 746}, 159 (2015).

\bibitem{MGinv} M. Giovannini, Phys. Rev.  D {\bf 108}, 123508 (2023).

\bibitem{MGinv2} M.~Giovannini, Eur. Phys. J. C {\bf 84},  67 (2024).

\bibitem{HT1} L. Boubekeur and D. H. Lyth, J. Cosmol. Astropart. Phys. {\bf 07}, 010 (2005).

\bibitem{HT2}  N. K. Stein and W. H. Kinney, J. Cosmol. Astropart. Phys. {\bf 03}, 027 (2023).

\bibitem{FR1} H. Motohashi and A. A. Starobinsky, J. Cosmol. Astropart. Phys. {\bf 11}, 025 (2019).

\bibitem{FR2}  M. Guerrero, D. Rubiera-Garcia, and D. Saez-Chillon Gomez, Phys. Rev. D {\bf 102}, 123528 (2020).

\bibitem{FR3} A. Mohammadi, T. Golanbari, S. Nasri, and K. Saaidi, Phys. Rev. D {\bf 101}, 123537 (2020).

\bibitem{RELGRAV} M.~Giovannini, Prog. Part. Nucl. Phys. \textbf{112}, 103774 (2020).

 \bibitem{BB1} R.~Brandenberger, JCAP {\bf 11}, 019 (2023).

\bibitem{BB2} V.~A.~Rubakov,  Phys.\ Usp.\  {\bf 57}, 128 (2014);   [Usp.\ Fiz.\ Nauk {\bf 184},  137 (2014)].

\bibitem{BB3} A.~Ashtekar and P.~Singh,  Class.\ Quant.\ Grav.\  {\bf 28}, 213001 (2011).

\bibitem{DD1} A.~Bhardwaj, E.~J.~Copeland and J.~Louko, Phys. Rev. D \textbf{99},  063520 (2019).

\bibitem{DD2} B.~Bonga and B.~Gupt, Gen. Rel. Grav. \textbf{48},  71 (2016).

\bibitem{BB4} M. Novello and S. E. P. Bergliaffa, Phys. Rept. {\bf 463}, 127 (2008).

\bibitem{OR1} C. Tolman, Phys. Rev. {\bf 38}, 1758 (1931).

\bibitem{OR2} G. Lema\^itre, Ann. Soc. Sci. Bruxelles A {\bf 53}, 51 (1933).

\bibitem{OR3} L.~Parker and S.~A.~Fulling,  Phys.\ Rev.\ D {\bf 7}, 2357 (1973).

\bibitem{OR4} A. A. Starobinsky, Sov.\ Astron.\ Lett. {\bf 4}, 82 (1978) [Pis'ma Astron. Zh. {\bf 4}, 155 (1978)].

\bibitem{OR5}  H. Nariai and K. Tomita, Progr. Theoret. Phys. {\bf 46}, 433 (1971). 

\bibitem{LL}  S. R. de Groot, V. A. van Leeuwen and Ch. G. van Weert , {\it Relativistic Kinetic Theory} (North-Holland, Amsterdam 1980).

\bibitem{LL1} L. D. Landau and E. M. Lifshitz {\it The Classical Theory of Fields} (New York, Pergamon Press, 1971).

\bibitem{EE} S. Weinberg, {\it Gravitation and Cosmology} ( Wiley New York, 1972).

\bibitem{WAD1} S.~Weinberg, Phys. Rev. D \textbf{67}, 123504 (2003).

\bibitem{WAD2} S.~Weinberg, Phys. Rev. D \textbf{70}, 083522 (2004).

\bibitem{WAD3} M. Giovannini, {\it A Primer on the Physics of the Cosmic Microwave Background}, 
(World Scientific, Singapore, 2008).

\bibitem{BBR1} J. Bardeen, Phys. Rev. D {\bf 22}, 1882 (1980).

\bibitem{SWL} J. M. Stewart and M. Walker, Proc. Roy. Soc. Lond. A {\bf 341}, 49 (1974).

\bibitem{BBR2} V.~N.~Lukash,  Sov.\ Phys.\ JETP {\bf 52}, 807 (1980) [Zh. Eksp. Teor. Fiz. {\bf 79}, 1601 (1980)].

\bibitem{BBR2b} V. Strokov, Astronomy Reports {\bf 51}, 431 (2007).

\bibitem{BBR2c} V. N. Lukash and I. D. Novikov, in {\em Observational and Physical Cosmology, II Canary Islands Winter School of Astrophysics}, 
edited by F. Sanchez, M. Collados, and R. Rebolo (Cambridge University Press, Cambridge, UK, 1992), p. 3.

\bibitem{BBR3} D.~H.~Lyth,  Phys.\ Rev.\ D {\bf 31}, 1792 (1985).

\bibitem{BBR4} M. Sasaki, Prog. Teor. Phys. {\bf 76}, 1036 (1986).

\bibitem{GI1} M.~Giovannini, Phys. Rev. D \textbf{95},  083506 (2017).

\bibitem{GI2} M.~Giovannini, Phys.\ Rev.\ D {\bf 70}, 103509 (2004).

\bibitem{GI3} M.~Giovannini, Class.\ Quant.\ Grav.\  {\bf 21}, 4209 (2004).

\bibitem{GI4} M.~Gasperini, M.~Giovannini and G.~Veneziano, Nucl.\ Phys.\ B {\bf 694}, 206 (2004).

\bibitem{nad3} M.~Giovannini,  Class.\ Quant.\ Grav.\  {\bf 23}, 4991 (2006).

\bibitem{nad1} K.~Enqvist, H.~Kurki-Suonio and J.~Valiviita, Phys.\ Rev.\  D {\bf 62}, 103003 (2000).

\bibitem{nad1a}  J.~Valiviita and V.~Muhonen,  Phys.\ Rev.\ Lett.\  {\bf 91}, 131302 (2003).

\bibitem{nad2} H.~Kurki-Suonio, V.~Muhonen and J.~Valiviita, Phys.\ Rev.\  D {\bf 71}, 063005 (2005).

\bibitem{nad2a} R.~Keskitalo, H.~Kurki-Suonio, V.~Muhonen and J.~Valiviita, JCAP {\bf 0709}, 008 (2007).


\end{thebibliography}
\end{document}